\documentclass[fleqn,usenatbib]{mnras}

\usepackage{newtxtext,newtxmath}
\usepackage[T1]{fontenc}

\DeclareRobustCommand{\VAN}[3]{#2}
\let\VANthebibliography\thebibliography
\def\thebibliography{\DeclareRobustCommand{\VAN}[3]{##3}\VANthebibliography}

\usepackage{graphicx}	% Including figure files
\usepackage{amsmath}	% Advanced maths commands
\usepackage{siunitx}
\DeclareSIUnit\erg{erg}
\DeclareSIUnit\bethe{B}
\usepackage{pdflscape}
\usepackage[inkscapelatex=false]{svg}
\usepackage{subfig}
\usepackage{orcidlink}
\title[$\nu$-driven CCSN Yields in GCE]{Neutrino-driven Core-collapse Supernova Yields in Galactic Chemical Evolution}

\author[Jost, Molero, Nav\'o, Arcones, Obergaulinger, Matteucci]{
Finia P. Jost\, \orcidlink{0009-0009-4357-0373}, $^{1}$\thanks{E-mail: finia.jost@tu-darmstadt.de}
Marta Molero\, \orcidlink{0000-0002-8854-6547}, $^{1}$\thanks{E-mail: marta.molero@tu-darmstadt.de}
Gerard Nav\'o\, \orcidlink{0000-0002-8984-0211}, $^{2}$
Almudena Arcones\, \orcidlink{0000-0002-6995-3032}, $^{1,3,4}$
\newauthor
\, Martin Obergaulinger\, \orcidlink{0000-0001-5664-1382
}, $^{2}$ and Francesca Matteucci\, \orcidlink{0000-0001-7067-2302} $^{5,6}$
\\
$^{1}$Institut für Kernphysik, Technische Universität Darmstadt, Schlossgartenstr. 2, Darmstadt 64289, Germany\\
$^{2}$Departament d’Astronomia i Astrofísica, Universitat de Val\`encia,
Edifici d'Investigaci\'o Jeroni Munyoz, C/ Dr. Moliner, 50, E-46100 Burjassot (Val\`encia), Spain\\
$^{3}$Max-Planck-Institut für Kernphysik, Saupfercheckweg 1, 69117 Heidelberg, Germany\\
$^{4}$GSI Helmholtzzentrum für Schwerionenforschung GmbH, Planckstr. 1, Darmstadt 64291, Germany\\
$^{5}$INAF, Osservatorio Astronomico di Trieste, Via Tiepolo 11, I-34131 Trieste, Italy\\
$^{6}$Dipartimento di Fisica, Sezione di Astronomia, Universit\`a degli studi di Trieste, Via G.B. Tiepolo 11, I-34143 Trieste, Italy
}

\date{Accepted XXX. Received YYY; in original form ZZZ}

\pubyear{2024}

\begin{document}
\label{firstpage}
\pagerange{\pageref{firstpage}--\pageref{lastpage}}
\maketitle

\begin{abstract}
We provide yields from 189 neutrino-driven core-collapse supernova (CCSN) simulations covering zero-age main sequence masses between $11$ and $75\ \mathrm{M}_\odot$ and three different metallicities. Our CCSN simulations have two main advantages compared to previous methods used for applications in Galactic chemical evolution (GCE). Firstly, the mass cut between remnant and ejecta evolves naturally. Secondly, the neutrino luminosities and thus the electron fraction are not modified. Both is key to obtain an accurate nucleosynthesis. We follow the composition with an in-situ nuclear reaction network including the 16 most abundant isotopes and use the yields as input in a GCE model of the Milky Way. We adopt a GCE which takes into account infall of gas as well as nucleosynthesis from a large variety of stellar sources. The GCE model is calibrated to reproduce the main features of the solar vicinity. For the CCSN models, we use different calibrations and propagate the uncertainty. We find a big impact of the CCSN yields on our GCE predictions. We compare the abundance ratios of C, O, Ne, Mg, Si, S, Ar, Ca, Ti, and Cr with respect to Fe to an observational data set as homogeneous as possible. From this, we conclude that at least half of the massive stars have to explode to match the observed abundance ratios. If the explosions are too energetic, the high amount of iron will suppress the abundance ratios. With this, we demonstrate how GCE models can be used to constrain the evolution and death of massive stars.
\end{abstract}


%%%%%%%%%%%%%%%%%%%%%%%%%%%%%%%%%%%%%%%%%%%%%%%%%%

%%%%%%%%%%%%%%%%% BODY OF PAPER %%%%%%%%%%%%%%%%%%

\section{Introduction}

 When massive stars explode as core-collapse supernovae (CCSNe) at the end of their evolution, they expel a significant fraction of their mass into the interstellar medium (ISM), including elements synthesized during hydrostatic burning phases as well as heavier elements beyond iron produced in explosive nucleosynthesis. As was shown by \citet{Romano2010QuantifyingUncertaintiesChemical}, the amount and composition of the ejecta are a key uncertainty in models of Galactic chemical evolution (GCE) (see also \citealp{Prantzos2018, Kobayashi2020OriginElementsCarbon, Palla2020}).
 %\citep{Kobayashi2020OriginElementsCarbon, Molero2023OriginNeutroncaptureElements}.
 These models require a large grid of stellar progenitors, varying in mass and metallicity. Even though the currently available yield grids come from updated stellar evolution and nucleosynthesis calculations, they can still significantly differ from each other. As a consequence, the same GCE models with the same assumptions on the main ingredients, produce different elemental abundances for different sets of yields. Self-consistent three-dimensional (3D) hydrodynamics simulations of CCSNe are now accessible (see e.g. \citealp{Janka2016PhysicsCoreCollapseSupernovae, OConnor2018ExploringFundamentallyThreedimensional, Ott2018ProgenitorDependenceCorecollapse, Burrows2020OverarchingFrameworkCorecollapse, Sandoval2021ThreedimensionalCorecollapseSupernova, Nakamura2022ThreedimensionalSimulationCorecollapse}) and are being used for explosive nucleosynthesis calculations \citep{Wang2024NucleosyntheticAnalysisThreeDimensional}. While the CCSN community is moving closer to the goal of a complete stellar yield set coming from these sophisticated models, simulating hundreds of CCSN models for several seconds still comes with a large computational cost and one-dimensional (1D) simulations remain to be useful tools for broad parameter studies. The stellar yields that have been used so far for GCE have been obtained from semi-analytic prescriptions \citep{Pignatari2016NUGRIDSTELLARDATA} or from CCSN simulations under the assumption of spherical symmetry. The lack of multi-dimensional effects prevents these models from exploding self-consistently.
 
 There have been many approaches to artificially trigger explosions by adding an artificial energy source to mimic the global effect of turbulence and convection. Classic approaches are the so-called piston method \citep{Woosley2007NucleosynthesisRemnantsMassive}, where a mass element is followed along a ballistic trajectory, or a kinetic bomb \citep{Limongi2018PresupernovaEvolutionExplosive}, where kinetic energy is placed in a chosen mass shell. In the thermal bomb approach \citep{Thielemann1996CoreCollapseSupernovaeTheira, Umeda2005VariationsAbundancePattern,  Imasheva2022ParametrizationsThermalBomb}, it is thermal energy that is injected. These methods are useful to obtain explosions and to investigate the dynamics and outcome of explosions of varying strength. However, these approaches have three unknowns: First, the mass cut between the proto-neutron star (PNS) and the ejecta is a free parameter in these methods and its position significantly influences the ejecta mass and composition. Second, observed explosion energies (e.g. from SN1987A) are used to calibrate the method, but the realistic explosion energy that should be reached by a particular progenitor model is not known. Third, the evolution of the composition in the innermost layers during collapse and explosion is not followed. Neutrino reactions change the electron fraction (due to neutrino reactions determining the proton to neutron ratio) in this region, which influences the explosive nucleosynthesis processes.
 
More recent approaches enhance the neutrino energy deposition in the gain layer in different ways. In the neutrino light bulb method \citep{Ugliano2012ProgenitorexplosionConnectionRemnant, Yamamoto2013PostshockrevivalEvolutionNeutrinoheatinga}, the interior of the PNS is replaced by inner boundary conditions informed by an analytic PNS core-cooling model. \citet{Ertl2016TwoparameterCriterionClassifying} and \citet{Sukhbold2016CorecollapseSupernovae120} found parameters of this method (P-HOTB) that reproduce properties of the Crab SN as well as SN1987A and used it to obtain predictions for explosion energies and Nickel production. The simulations were mapped to a piston model to calculate the nucleosynthesis. The drawback of this approach is that the neutrino interactions happening in the innermost layers of the ejecta, which change the thermodynamic properties and the composition, are not captured during the explosion phase. This issue has been addressed with the so-called PUSH method \citep{Perego2015PUSHingCorecollapseSupernovae, Ebinger2019PUSHingCorecollapseSupernovae, Curtis2018PUSHingCorecollapseSupernovae,Ebinger2020PUSHingCorecollapseSupernovae, Ghosh2022PUSHingCorecollapseSupernovae}. There, a fraction of the energy carried by muon and tau neutrinos is deposited in the gain region behind the shock. This additional neutrino energy deposition is parameterized in dependence of the compactness parameter of the stellar progenitor (introduced in Section \ref{models}). With this, the explosion energy and Ni ejecta of several observed supernovae including SN1987A and the Crab can be reproduced. The PUSH method allows for a neutrino-driven explosion mechanism with enhanced energy deposition without modifying the electron-flavor neutrino luminosities. Additionally, the evolution of the PNS is simulated self-consistently and the mass cut between remnant and ejecta emerges naturally. The same is true for an alternative approach called 'Simulating Turbulence In Reduced dimensionality' (STIR), introduced in \citet{Couch2020SimulatingTurbulenceaidedNeutrinodriven}, which is more closely guided by turbulent features (the typical turbulent speed velocity) of 3D simulations. The strength of convection and turbulence is estimated using a modified time-dependent mixing length theory and included in the fluid equations.\\

In this study, we achieve our artificial explosions by enhancing the neutrino heating in the gain layer similar to the PUSH method. Instead of introducing an absorption term that depends on the heavy-flavor neutrinos, we multiply the energy gain from  neutrino reactions (mostly charged-current reactions with electron-flavor neutrinos), by a 'heating factor'. This multiplication factor is restricted to the gain region and thus only enhances the neutrino heating, not the cooling. A similar approach has been introduced in \citet{OConnor2010NewOpensourceCode} and \citet{Witt2021PostexplosionEvolutionCorecollapse}. Note that our method shares with the PUSH and STIR methods the following advantages. First, the PNS is included in the simulation domain and the mass cut evolves naturally in the simulation. Second, we only influence the total energy of the system, so the neutrino transport is not modified. Our CCSN simulations include a state-of-the-art, two-moment neutrino treatment \citep{Just2015NewMultidimensionalEnergydependent}, which takes all important neutrino interactions into account and allows for a self-consistent determination of the electron fraction in the ejecta. The nucleosynthesis in the innermost ejecta depends directly on the explosion dynamics, and specifically, on the electron fraction in these layers.
\\

 All approaches to explode massive stars in spherical symmetry have one or more free parameters (e.g., the amount of energy injected, where it is injected, the mass cut, the heating factor), which have to be calibrated to either more realistic 3D simulations or to properties of observed CCSNe. The choice of the calibration significantly impacts the final yields and can even decide between successful and failed explosion for the same stellar progenitor (see e.g. \citealp{Young2007UncertaintiesSupernovaYields, Couch2020SimulatingTurbulenceaidedNeutrinodriven}). The question of explodability is a key problem in CCSN theory since it was found that there is no simple relation between the zero-age main sequence mass of a star and its final fate (see e.g. \citealp{Ugliano2012ProgenitorexplosionConnectionRemnant, Sukhbold2016CorecollapseSupernovae120, Muller2016SimpleApproachSupernova}). There is no agreement between different studies on which stellar progenitors are expected to explode, although several criteria have been developed to enable this prediction \citep{OConnor2011BlackHoleFormation, Ertl2016TwoparameterCriterionClassifying, Gogilashvili2022ForceExplosionCondition, Wang2022EssentialCharacterNeutrino, Boccioli2023ExplosionMechanismCorecollapse}. Additional uncertainties that enter in CCSN simulations are related to the adopted high-density equation of state (see e.g. \citealp{Schneider2019EquationStateEffects, Yasin2020EquationStateEffects, Pascal2022ProtoneutronStarEvolution}), neutrino transport (see e.g. \citealp{Iwakami2020SimulationsEarlyPostbounce, Wang2023EffectsDifferentClosure}), pre-supernova stellar models (see e.g. \citealp{Woosley2002EvolutionExplosionMassive, Ekstrom2012GridsStellarModels, Muller2017SupernovaSimulations3D, Cristini20173DHydrodynamicSimulations, Limongi2018PresupernovaEvolutionExplosive, OConnor2018ExploringFundamentallyThreedimensional,Fields2021ThreedimensionalHydrodynamicSimulations, Yoshida2021ThreedimensionalHydrodynamicsSimulations, Vartanyan2022CollapseThreedimensionalExplosion}), the impact of magnetic fields and rotation (see e.g. \citealp{Kuroda2020MagnetorotationalExplosionMassive, Jardine2022GravitationalWaveSignals, Obergaulinger2022MagnetorotationalCoreCollapse, Varma20223DSimulationsStrongly, Li2023EffectsRotationMetallicity}), as well as the nucleosynthesis (see e.g. \citealp{Harris2017ImplicationsPostprocessingNucleosynthesis, Navo2023CorecollapseSupernovaSimulations, Sieverding2023TracerParticlesCorecollapse, Wang2024NucleosyntheticAnalysisThreeDimensional}). In this study, we use more than one calibration to account for this uncertainty and show how the uncertainty propagates into GCE models. Our aim is not to provide exact results, but to observe trends by including extreme scenarios of all stars exploding or only half of them.
 \\
 
 Besides observations of the light curves and spectra of individual events, GCE models build the bridge to compare the yields predicted by CCSN simulations with the observed chemical properties of the ISM and stellar populations they ultimately try to reproduce. With the growing amount of observational data that is arriving from several Galactic archaeology surveys (e.g., the Galactic Archaeology with HERMES survey, GALAH, \citealp{Buder2021}; the Apache Point Observatory Galactic Evolution Experiment project, APOGEE, \citealp{Majewski2017}; the AMBRE project, \citealp{delaverny2013}; the Gaia-ESO project, \citealp{Mikolatisi2014}), the need to build a self-consistent model of GCE able to provide stringent constraints on both stellar astrophysics and on the evolutionary history of our own Galaxy is becoming more and more important. Interpretation of observations from abundance surveys may be complex and, often, chemical evolution studies have to rely on a certain number of assumptions due to our limited understanding of some key parameters (e.g., star formation, initial mass function, gas flows and stellar migration). Nevertheless, GCE models have been proven to be powerful tools in investigating the production sources and timescales of different isotopes. To investigate the origin of the elements it is of great importance to build a self-consistent chain between stellar evolution, nucleosynthesis calculations and, finally, GCE models. In this study, we go all the way from simulating the collapse and explosion of massive stars, providing the nucleosynthesis results and using them as input for a GCE model to be able in the end to compare the results to observations of abundances for our own Galaxy.
 \\
 
 The paper is structured as follows. In Section \ref{sec: core-collapse supernova yields}, we introduce the setup and input for our CCSN simulations, discuss the relation of progenitor properties and explosion dynamics, and explain how we obtain our massive star yields for GCE. In Section \ref{gce}, we describe the adopted GCE and our main results for the solar abundances, the metallicity distribution function of the solar neighborhood and, finally, the abundance patterns of the elements produced by our new CCSN simulations. Section \ref{sec: conclusions} summarises and discusses the results.

\section{Core-collapse supernova yields}
\label{sec: core-collapse supernova yields}

\subsection{Hydrodynamic code} \label{code}
We have simulated 189 models of CCSNe with the magneto-hydrodynamics code \textsc{Aenus-Alcar} \citep{Just2015NewMultidimensionalEnergydependent,Obergaulinger2022MagnetorotationalCoreCollapse}. The code includes special relativity, a pseudo-relativistic gravitational potential and a two-moment (M1) neutrino transport. The simulations employ spherical symmetry and a logarithmic grid with $\SI{1000}{}$ cells in the radial direction and a central width of $\SI{200}{\metre}$. The simulations end after $1.5-5$ seconds once the yields have saturated. The SFHo nuclear equation of state (EOS) \citep{Steiner2013CorecollapseSupernovaEquations} describes the matter at $T>\SI{5.8}{\giga\kelvin}$ assuming a nuclear statistical equilibrium (NSE) between protons, neutrons, alpha particles, and a representative heavy nucleus. Below this threshold NSE freezes out and the composition is obtained with a reduced nuclear reaction network \citep{Navo2023CorecollapseSupernovaSimulations} and an EOS taking into account photons, electrons, positrons and ions \citep{Timmes1999AccuracyConsistencySpeed}. In this work, we employ the 16-$\alpha$ chain reaction network, which has the smallest computational cost out of the implemented networks and gives comparable results to the larger networks in 1D simulations (for a comparison, see \citealp{Navo2023CorecollapseSupernovaSimulations}). Between those regimes, the two EOS are interpolated linearly in order to prevent discontinuities in the thermodynamic quantities. The initial conditions for the reduced nuclear reaction network are given by an NSE solver considering the same nuclei as the network at $T=\SI{5.8}{\giga\kelvin}$.\\

As explained in the introduction, we artificially enhance the neutrino heating in the gain layer with a 'heating factor' (HF). This is a multiplication factor to the energy source term $Q_\text{E}$, which describes the change of the energy density owing to neutrino reactions in the corresponding hydrodynamic equation. The HF is applied when two conditions are met: First, the density is required to be below a threshold ($\rho < \SI{1e10}{\gram\per\cubic\centi\metre}$) to confine the region to be outside the PNS. Second, there must be a positive energy gain from neutrino-matter interactions ($Q_\text{E} > 0$) in the cell before adding the artificial heating. These conditions confine the area, where the HF is applied, to the gain region, so neutrino heating is enhanced while neutrino cooling is not modified. Since the energy increase via the HF is artificial, it is desirable to reduce it to the necessary minimum for a successful explosion. Therefore, the HF is applied only until the shock has reached a certain radius, after which it is unlikely to fall back onto the PNS. This radius is the second free parameter of our model. Once the shock reaches the chosen radius, the HF is set to one and has no further impact. It has been tested that this abrupt switch-off does not lead to instabilities in the hydrodynamic quantities and that a linear transition produces the exact same results.

\subsection{Stellar models} \label{models}
\begin{figure}
	\includegraphics[width=\columnwidth]{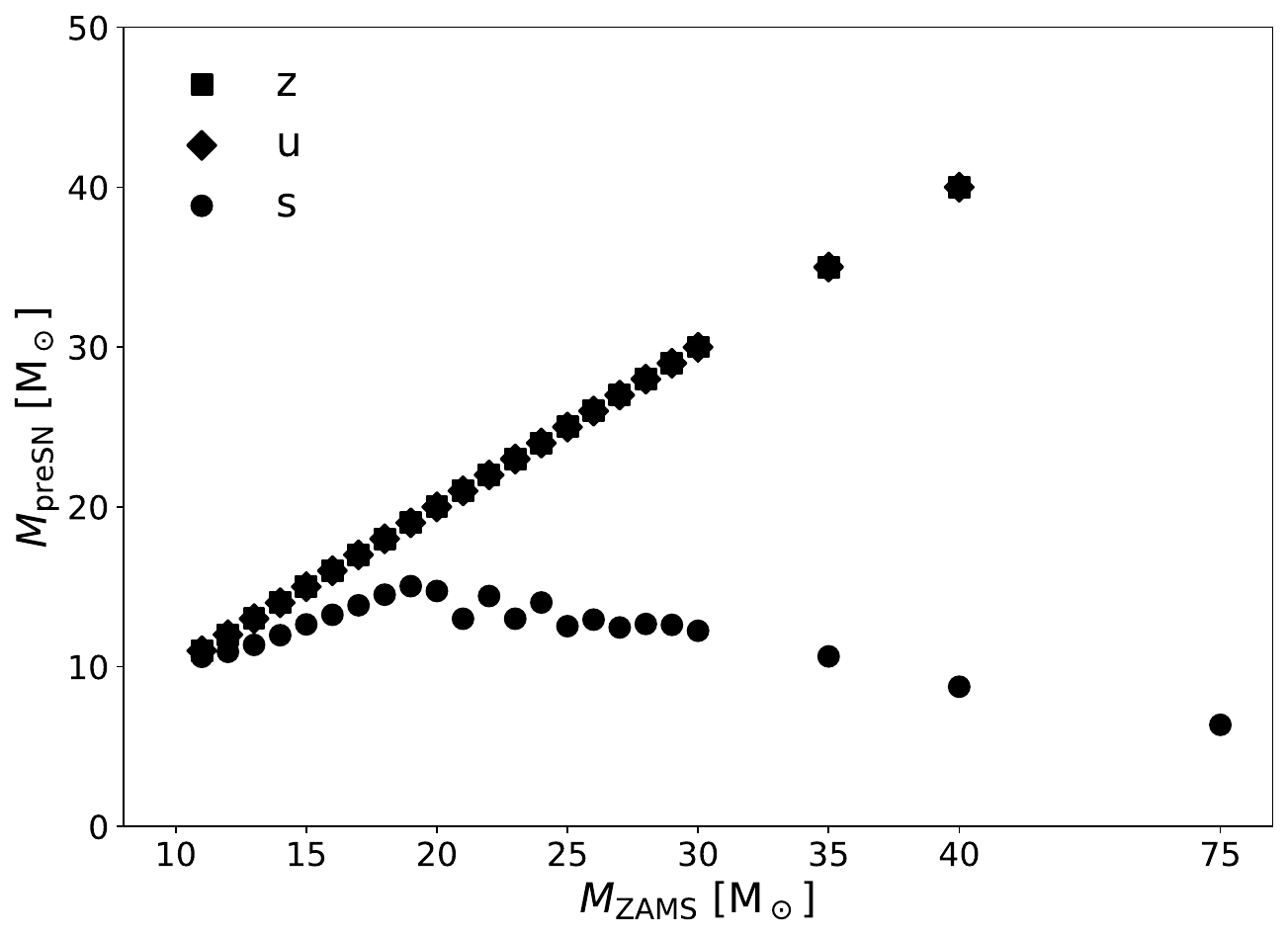}
    \caption{Progenitor mass at the onset of our simulations for all stellar models employed from the series of \citet{Woosley2002EvolutionExplosionMassive}. The symbols indicate the initial metallicity.}
    \label{fig:Mpresn}
\end{figure}

\begin{figure}
	\includegraphics[width=\columnwidth]{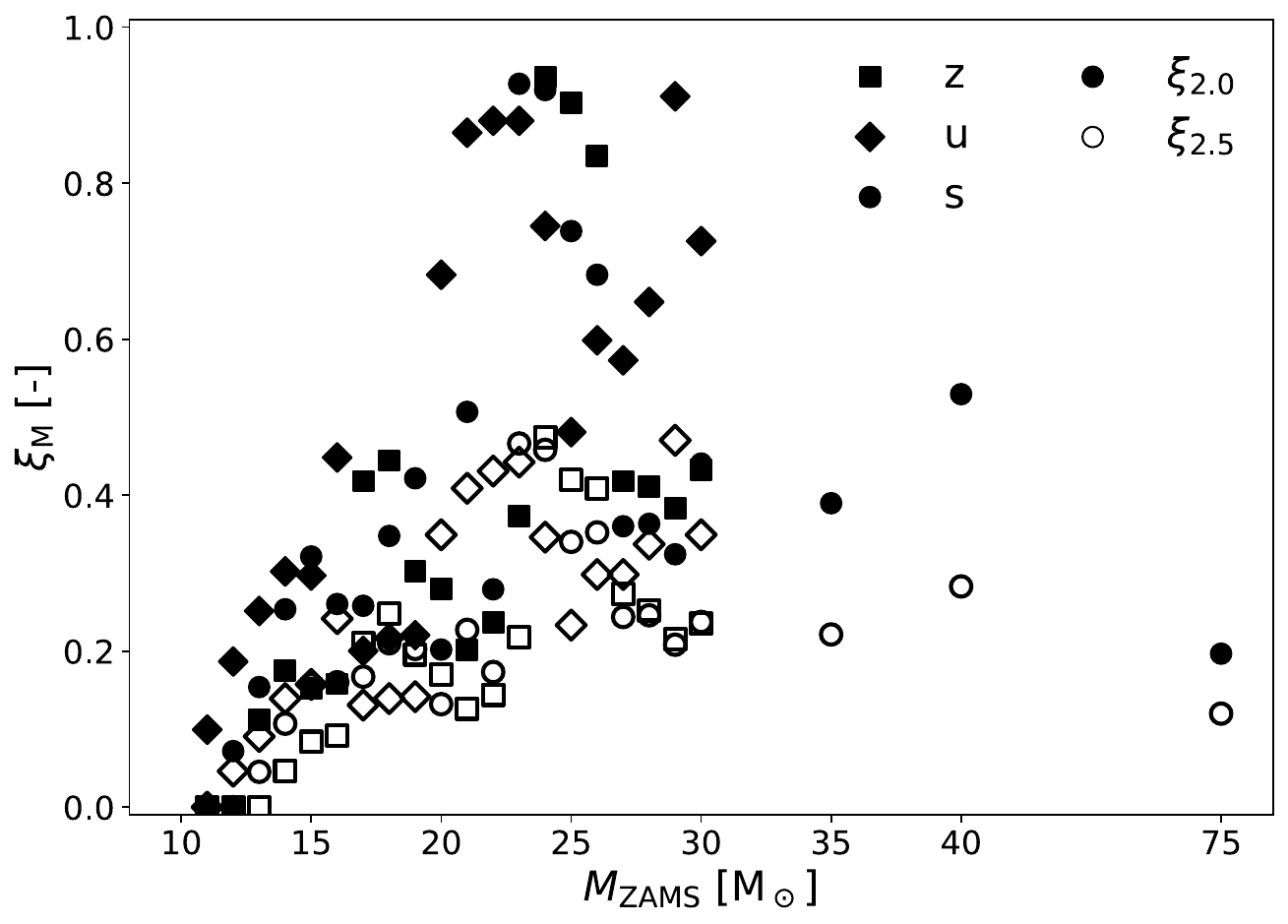}
    \caption{Compactness parameter for all progenitor models for mass shells $M=\SI{2.0} {}M_\odot$ (filled circles) and $M=\SI{2.5}{}M_\odot$ (open circles). If it is zero, less than $2\ M_\odot$ lie within the simulation region. The pattern is the same, but it is more pronounced for the mass shell enclosing $2\ M_\odot$. The symbols indicate the initial metallicity.}
    \label{fig:compactness}
\end{figure}

For a systematic study on the impact of the progenitor mass and structure on the explosion outcome, we run simulations with in total $63$ progenitor models of the non-rotating series of \citet{Woosley2002EvolutionExplosionMassive} (WHW02) with zero-age main sequence (ZAMS) masses of $11-30\ M_\odot$ in steps of $1\ M_\odot$ for three different metallicities, namely primordial, $10^{-4}$ solar, and solar metallicity. Additionally, we simulate the explosion of the solar metallicity models with ZAMS masses $35$, $40$, and $75\ M_\odot$. The progenitors with the same masses, but different metallicity likely form black holes, which cannot be captured in our simulations. Note that the stellar models with solar metallicity experience a significant amount of mass loss during their evolution and their mass will be much reduced at the time of the collapse, where our simulations begin (see Fig.~\ref{fig:Mpresn}). This is not the case for progenitors of primordial metallicity and also not for the intermediate metallicity of $10^{-4}$ solar. It is known that stars can be fast rotating and that this affects the structure and composition of the star (for a detailed discussion, see \citealp{Limongi2018PresupernovaEvolutionExplosive}). Nevertheless, we do not include any rotating stars in this study. We denote the stellar models as 'mX', where 'm' indicates the metallicity ('s' stands for solar metallicity, 'u' for $10^{-4}$ solar, and 'z' for primordial metallicity) and 'X' is the ZAMS mass in solar masses. The simulations are denoted as 'mX-HF', where the HF is added to the previous notation. \citet{OConnor2011BlackHoleFormation} define the compactness parameter
\begin{equation}
    \xi_\text{M}=\frac{M/M_\odot}{R(M_{\text{bary}}=M)/\SI{1000}{\kilo\metre}}|_{t=t_{\text{bounce}}}
\end{equation} 
to link the progenitor structure to the explosion outcome in a simple way. They state that there is a threshold at which higher compactness always leads to black hole formation. The compactness is evaluated at the time of bounce as this is the most unambiguous point in core collapse. In Fig.~\ref{fig:compactness}, we display the compactness parameter evaluated for two mass shells for our set of progenitors. We come back to the relation of progenitor structure and its final fate in Section \ref{explosion}.

\subsection{Calibration} \label{calibration}
We employ several calibrations of our setup to estimate the uncertainty of our CCSN yields and propagate this uncertainty to the model of GCE. First, we calibrate to SN1987A, which is the most studied observed supernova event and has rather well constrained properties, in particular the explosion energy $ E_{\text{exp}}~=~\SI{1.1\pm 0.3}{\times 10^{51}erg}~=~\SI{1.1\pm 0.3}{B}$ \citep{Blinnikov2000RadiationHydrodynamicsSN} and the amount of ejected Nickel $M(^{56}\text{Ni})~=~\SI{0.071\pm 0.003}{}\ M_\odot $ \citep{Seitenzahl2014LightCurveSN1987A}. Four models of \citet{Woosley2002EvolutionExplosionMassive} have been used for the calibration: the solar metallicity models with ZAMS masses of $18\ M_\odot$, $18.8\ M_\odot$ and $19.8\ M_\odot$ as well as the primordial-metallicity model with a ZAMS mass of $18\ M_\odot$ (see \citealp{Ugliano2012ProgenitorexplosionConnectionRemnant, Perego2015PUSHingCorecollapseSupernovae, Sukhbold2016CorecollapseSupernovae120, Ebinger2019PUSHingCorecollapseSupernovae, Curtis2018PUSHingCorecollapseSupernovae} for a similar calibration). For the sake of reaching the high explosion energy of the order of $\SI{1}{\bethe}$, a very high HF of $5.6$, which is kept until the shock reaches $\SI{12000}{\kilo\metre}$, is needed for these progenitors. These values represent the upper limits for our free parameters. With this setup, all stellar models explode. On the lower end, we calibrate to a 3D simulation of the solar metallicity model with a ZAMS mass of $25\ M_\odot$ without rotation nor magnetic field, and without HF as this is not needed in 3D models. The simulation led to an explosion energy of $E_{\text{exp}}=\SI{0.62}{\bethe}$, which can already be met with a HF of $2.4$, which is kept until the shock reaches $\SI{4000}{\kilo\metre}$. With these low values, only a fraction of about 40~per cent of our models explode. For stars that do not explode, all material falls back onto the remnant and no matter is ejected. We use these values as a lower limit. We find that for many models, the threshold of explodability lies in the region of the HF between $2.5$ and $2.7$ and decide to use $2.7$ as an intermediate and more standard case. Again, we keep the HF until the shock has reached $\SI{4000}{\kilo\metre}$. With this, we have nine failed explosions (about 14~per cent) in agreement with the inferred fraction of failed SNe of about 16~per cent \citep{Neustadt2021SearchFailedSupernovae}.

Note that it is not clear that the multiplication factor enhancing the neutrino energy deposition should be a universal number. It is likely that the strength with which turbulence and convection aid the neutrino-driven explosion varies with the stellar structure and resulting explosion dynamics. In the PUSH method, the multiplication factor of the energy deposition based on the heavy-flavor neutrino luminosity is parameterized in dependence to the compactness parameter of the stellar progenitor \citep{Ebinger2019PUSHingCorecollapseSupernovae}. This is motivated by the fact that higher explosion energies are obtained for progenitors around $20\ M_\odot$ and intermediate compactness $\xi_{2.0}\sim0.3$ (like SN1987A) than for lower-mass stars around $10-15\ M_\odot$ with low compactness $\xi_{2.0}\sim0.0$ (like the Crab, see e.g. \citealp{Temim2024DissectingCrabNebula}). They expect very compact stars $\xi_{2.0}\sim0.7$ to form black holes and not explode. With a compactness-dependent calibration, they can match the amount of Nickel that is ejected and the explosion energy of several observed CCSN events including the Crab and SN1987A. For this, the multiplication factor to the neutrino heating of the PUSH method has to be enhanced around $\xi_{2.0}\sim0.3$. While such a parametrization might be more adequate than adopting a constant factor to find the best match to observational constraints, it is not clear how exactly this should be done as the question which outcome is expected for a specific stellar progenitor is not settled. For example, we have used four progenitor models to calibrate to the explosion energy and Ni ejecta of SN1987A and they all needed $\text{HF}=5.6$ to reach $E_{\text{exp}}=\SI{1}{\bethe}$ (see Section~\ref{calibration}). In the same mass regime, there is model s19, which achieves an explosion energy of $E_{\text{exp}}=\SI{0.95}{\bethe}$ already with $\text{HF}=2.7$.

When using $\text{HF}=2.7$ for all models, the pattern of explosion energy with ZAMS mass that is expected from observations is roughly reproduced as can be seen in Fig.~\ref{fig:en}. On the lower-mass end, explosions are less energetic $E_{\text{exp}}=0.3-\SI{0.6}{\bethe}$. The explosion energies rise up with $E_{\text{exp}}=1-\SI{1.7}{\bethe}$ on the higher end between $M_{\text{ZAMS}}=15-25$ and decrease again towards higher masses. For models with lower mass, the agreement with the results of the PUSH method is better with HF = $5.6$. Above $20~M_\odot$, the results of the lower HFs are more similar. The high explosion energies achieved for models around $22-25~M_\odot$ are in agreement with results of the STIR approach (compare Fig.~\ref{fig:en} with Fig.~8 in \citealp{Couch2020SimulatingTurbulenceaidedNeutrinodriven}). This will be discussed further in the following Section. The large scatter in this pattern makes a more precise fit difficult. Due to the uncertainties involved in the choices of such a calibration, we provide three sets of simulations employing different values for our free parameters. We interpret the results as uncertainty estimates, which we propagate to the application in the model of GCE.

In 1D simulations, the total energy of unbound material saturates or even decreases after a few hundred milliseconds. Since there is no simultaneous accretion and ejection in a spherically symmetric model, no more energy is injected into matter that is unbound. In multi-dimensional models, the explosion energy keeps rising for several seconds due to continuing accretion on the PNS (see e.g. \citealp{Bruenn2016DevelopmentExplosionsAxisymmetric,Witt2021PostexplosionEvolutionCorecollapse,Burrows2024PhysicalCorrelationsPredictions}). Since the explosion energy is expected to increase further for several seconds, we use the peak value of the total energy of unbound material (without overburden) as diagnostic explosion energy instead of the value at the last time step. One has to keep in mind the limitations of 1D simulations and that they are more suitable to observe trends rather than exact numbers. We take this into account by varying our free parameters and exploring a large range of possible explosion outcomes.

\subsection{Explosion dynamics} \label{explosion}
We find a high dependency of the explosion outcome on the choice of the HF, even decisive about a successful explosion as it was explained in the previous Section. For many models, the threshold of explodability is around $\text{HF}=2.5-2.7$. In general, a higher HF adds more energy to the system, which results in earlier explosions of higher explosion energy. With the calibration that yields the highest explosion energies, all models explode and roughly at the same time around $\SI{0.2}{\second}$ post-bounce. But there are exceptions to this trend, because the dynamics of the explosion mechanism highly depend on the stellar structure. A lower HF leads to later explosions, so that the layers behind the shock spend more time in the vicinity of the PNS, where they can gain most energy. Additionally, the increased accretion in these models powers stronger neutrino luminosities. This can, in some cases, lead to a higher explosion energy with a lower HF (e.g. models z25 and u29).

Many studies aimed to infer the explosion outcome from characteristics of the progenitor structure (see e.g. \citealp{Sukhbold2016CorecollapseSupernovae120, Ertl2016TwoparameterCriterionClassifying, Muller2016SimpleApproachSupernova, OConnor2011BlackHoleFormation, Ebinger2019PUSHingCorecollapseSupernovae, Wang2022EssentialCharacterNeutrino}). For some time, they have mostly reproduced the statement of \citet{OConnor2011BlackHoleFormation} that a high compactness parameter has a tendency to correlate with black hole formation. \citet{Ebinger2019PUSHingCorecollapseSupernovae} adopted the assumption that progenitors with high compactness are unlikely to explode and calibrated the PUSH method in a compactness-dependent way, artificially reproducing this assumption. However, \citet{Couch2020SimulatingTurbulenceaidedNeutrinodriven} do not find that progenitors with a high compactness are particularly difficult to explode. They highlight that this is a distinguishing feature between the STIR method and the 1D neutrino-driven explosion model of \citet{Sukhbold2016CorecollapseSupernovae120} and \citet{Ertl2016TwoparameterCriterionClassifying}. Note that \citet{Couch2020SimulatingTurbulenceaidedNeutrinodriven} use the same set of progenitors that was used in \citet{Sukhbold2016CorecollapseSupernovae120}. They draw the conclusion that including the impact of turbulent convection in 1D simulations results in dynamics that are significantly different from those for purely neutrino-driven 1D explosions. Meanwhile, studies of multi-dimensional CCSN simulations have confirmed that stars with higher compactness are actually easier to explode. \citet{Vartanyan2023NeutrinoSignatures100} have analysed 100 2D simulations and they state that models between $12$ and $15\ M_\odot$ are more difficult to explode, because they have shallow Silicon-Oxygen (Si/O) interfaces, whose accretion is not sufficient to revive the shock expansion. In many progenitors, the accretion of the Si/O-interface is accompanied with a sudden decrease in density and thus a drop in the accretion ram pressure, which can aid the initiation of the explosion (see \citealp{Summa2016ProgenitordependentExplosionDynamics} and \citealp{Boccioli2023ExplosionMechanismCorecollapse} for more details on the role of the Si/O-interface). \citet{Burrows2024PhysicalCorrelationsPredictions} have analysed a suite of twenty state-of-the-art 3D CCSN simulations. Only four of their simulations result in black hole formation and they state that the models with ZAMS masses $12.25$ and $14\ M_\odot$ have a too low compactness to explode.

\begin{figure}
	\includegraphics[width=\columnwidth]{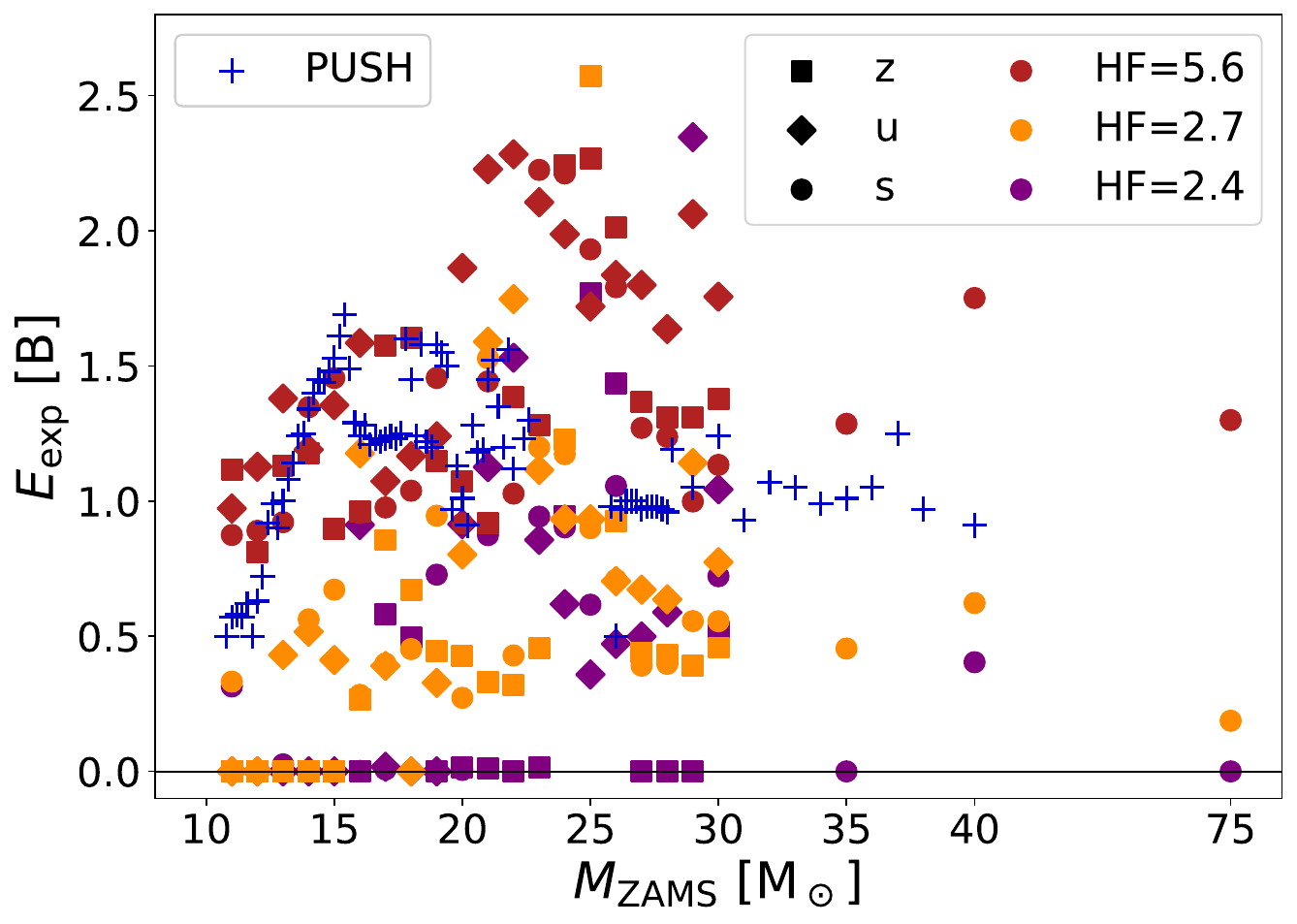}
    \caption{Diagnostic explosion energy in relation to ZAMS mass for all models. If $E_\mathrm{exp}$ is zero, the stellar model did not explode in our simulation. The different symbols indicate the initial metallicity, the colors the adopted heating factor. For comparison, we include the corresponding data from the PUSH approach \citep{Ebinger2019PUSHingCorecollapseSupernovae} as blue crosses. Compare also to Fig.~8 in \citet{Couch2020SimulatingTurbulenceaidedNeutrinodriven}.}
    \label{fig:en}
\end{figure}

\begin{figure}
	\includegraphics[width=\columnwidth]{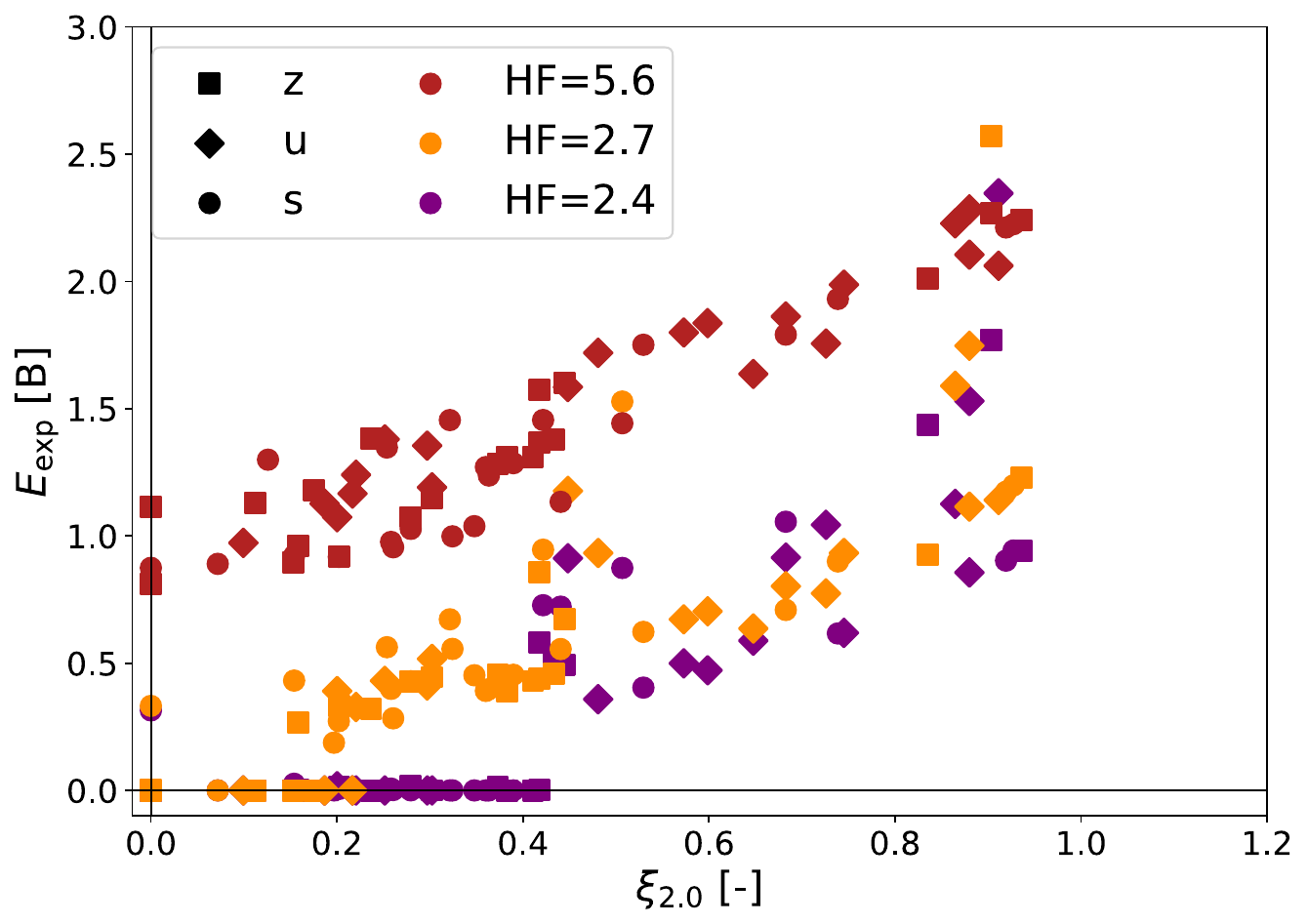}
    \caption{Relation of diagnostic explosion energy and compactness parameter $\xi_{2.0}$ for all models. If $\xi_{2.0}$ is zero, less than $2\ M_\odot$ lie within the simulation region. If $E_\mathrm{exp}$ is zero, the stellar model did not explode in our simulation. The different symbols indicate the initial metallicity, the colors the adopted heating factor.}
    \label{fig:compact_en}
\end{figure}

The present approach, which is purely neutrino-driven and spherically symmetric, also readily explodes progenitors with high compactness, and even with less additional heating than progenitors with a small compactness parameter. Our explodability pattern aligns well with the findings of multi-dimensional simulations and STIR (see Figs.~\ref{fig:en} and \ref{fig:compact_en}). Low compactness stars with $\xi_{2.0}<0.4$ fail to explode with the lowest HF. Around the threshold of $\xi_{2.0}=0.4$, there is no clear cut and stars with similar compactness can explode or not explode for the same HF, depending on their detailed structure. Above the threshold, the simulations result in explosion. The models that only explode with our highest HF mostly have ZAMS masses in the range of $11-15\ M_\odot$ (see zero explosion energy for non-exploding models in Fig.~\ref{fig:en}).

The compactness parameter does not only indicate whether a star is likely to explode, but is also connected to the explosion energy. \citet{Barker2022ConnectingLightCurves} employ the STIR method \citep{Couch2020SimulatingTurbulenceaidedNeutrinodriven} and find a positive correlation between the compactness parameter of a progenitor and its explosion energy and explain this with the higher gravitational binding energy that is available in the explosion of models with more compact cores. \citet{Burrows2024PhysicalCorrelationsPredictions} have confirmed this correlation with their suite of 3D simulations. Our simulations (though purely neutrino-driven and 1D) reproduce this correlation of explosion energy with progenitor compactness (see Fig.~\ref{fig:compact_en}), adopting a constant HF as explained in the previous Section~\ref{calibration}.\\

\subsection{Yields} \label{yields}
From our simulations, we obtain the CCSN yields with a reduced, in-situ nuclear reaction network. We have compared the abundance pattern of the in-situ network with a full network calculation in post-processing using WinNet \citep{Reichert2023NuclearReactionNetwork} for a subset of our simulations and we have found that the abundances of the main isotopes are well reproduced with the reduced network \citep{Navo2023CorecollapseSupernovaSimulations}. Note that it is the shell just above the PNS that produces significant amounts of $^{44}$Ti, $^{48}$Cr, $^{52}$Fe, $^{54}$Fe, and mostly $^{56}$Ni. Therefore, the exact location of the mass cut between the remnant and the ejecta determines the yields of a CCSN. Our simulations include the region of the PNS and do not require an assumption on the mass cut.

\begin{figure}
	\includegraphics[width=\columnwidth]{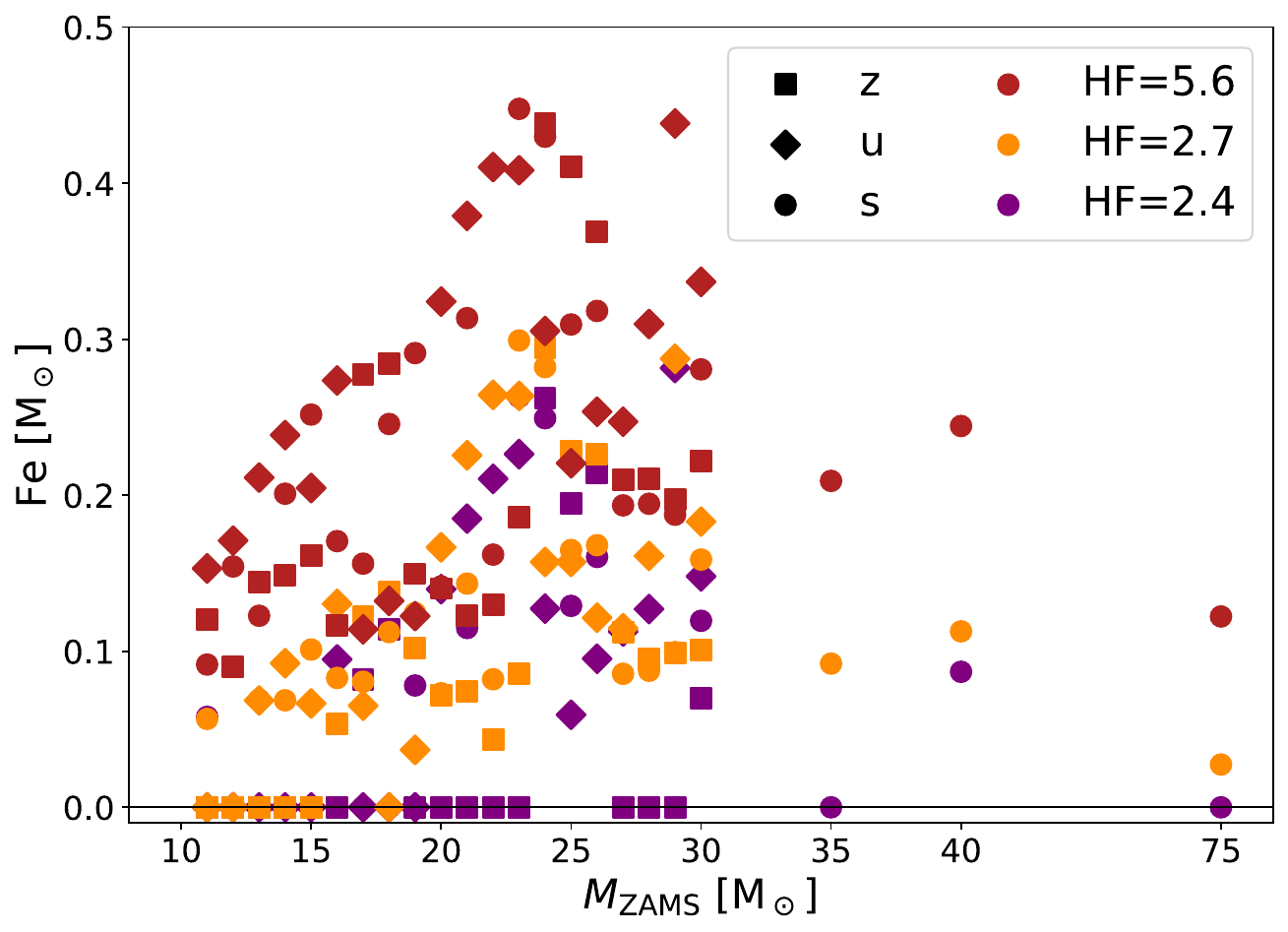}
    \caption{Iron ejected during the explosion (total, not net and no wind) for all models. The different symbols indicate the initial metallicity, the colors the adopted heating factor.}
    \label{fig:fe_m}
\end{figure}

\begin{figure}
	\includegraphics[width=\columnwidth]{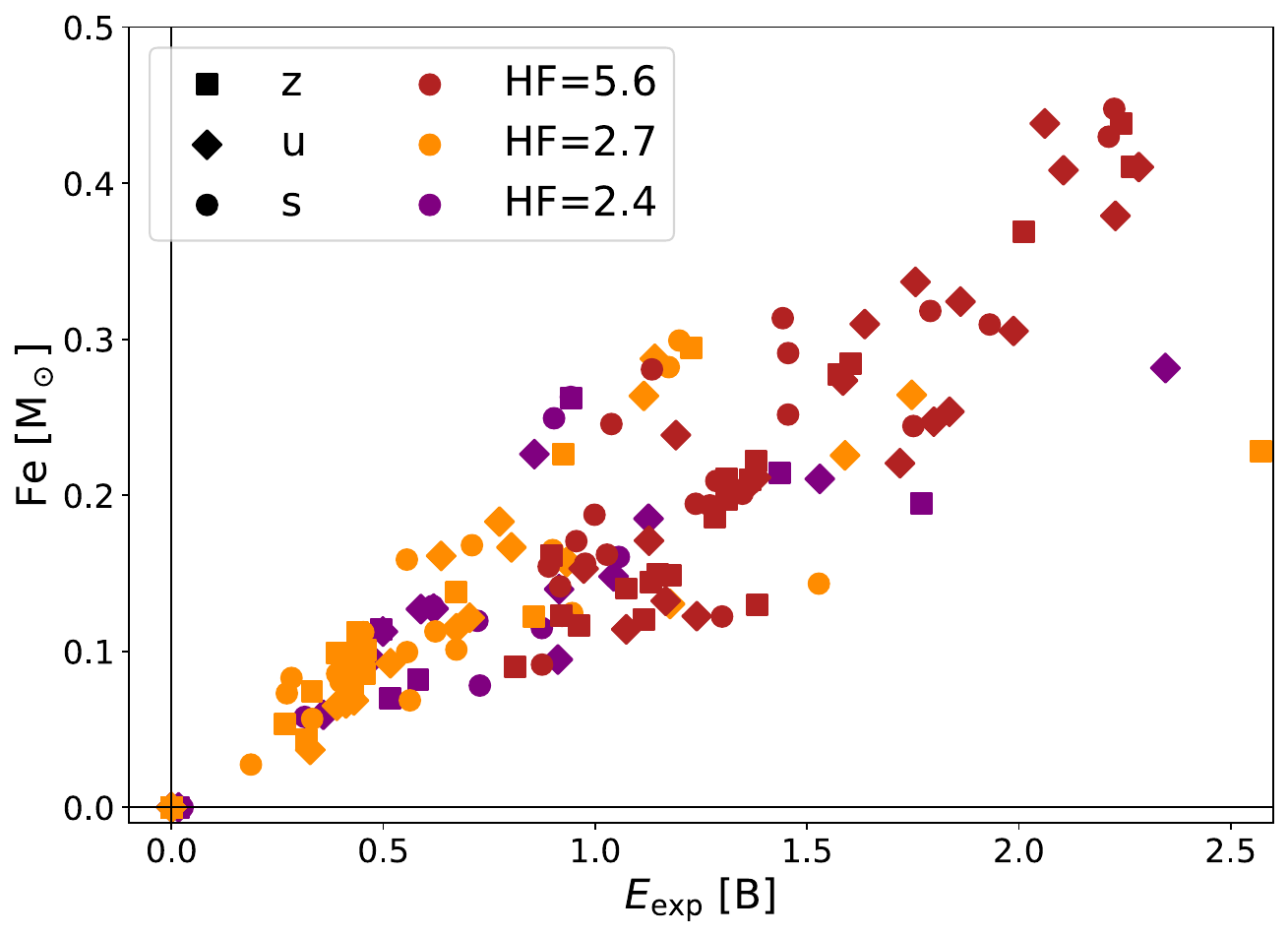}
    \caption{Relation of ejected iron and diagnostic explosion energy for all models. The different symbols indicate the initial metallicity, the colors the adopted heating factor.}
    \label{fig:fe_en}
\end{figure}

In Fig.~\ref{fig:fe_m}, we present the amount of iron that is ejected in all our CCSN simulations. When the iron yield is zero, that means that the simulation did not result in an explosion and correspondingly, all yields are zero. The iron comes from the decay of radioactive $^{56}$Ni. We find a correlation between the ejected amount of $^{56}$Ni and the explosion energy that is independent on the metallicity or adopted heating factor (see Fig.~\ref{fig:fe_en}). This is a well-known relation, which has been theoretically expected from semi-analytic CCSN models (e.g. \citealp{Pejcha2015LandscapeNeutrinoMechanism}) as well as from simulations (e.g. \citealp{Sukhbold2016CorecollapseSupernovae120}, \citealp{Ebinger2019PUSHingCorecollapseSupernovae}, \citealp{Burrows2024PhysicalCorrelationsPredictions}) and has been observed astronomically (e.g. \citealp{Hamuy2003ObservedPhysicalProperties}, \citealp{Muller2017NickelMassDistribution}). Comparing more quantitatively to the 3D simulations by \citet{Burrows2024PhysicalCorrelationsPredictions}, our results agree in that simulations with explosion energies around $0.5-\SI{1.25}{B}$ eject around $0.1\ M_\odot$ of $^{56}$Ni. For higher explosion energies, there is a better agreement with their results for our two lower heating factors. The amount of $^{56}$Ni ejected by our simulations with $\text{HF}=5.6$ is much higher than what they find for explosion energies between $1.5-\SI{2.5}{B}$.
\\

In the following, we describe how we obtain the massive star yields from our CCSN simulations to apply them in our GCE model. In our GCE model, we take away the ZAMS mass of a star from the ISM when it is formed and give back the same initial mass with its initial composition to the ISM when the star dies. The stellar nucleosynthesis, both hydrostatic and explosive, is taken into account by adding net yields to the initial ones when they are given back. \\
We obtain the net yields in the following way. Material in cells that lie above the innermost cell with positive total energy density and that have positive radial velocity are considered to be unbound. For each of the 16 isotopes in our network, the mass of all unbound cells is summed up, once the yields have saturated (after $1$-$\SI{5}{\second}$, depending on the model). We add, to the mass that is already unbound, all mass that is still outside the shock radius with the composition of the progenitor prior to the supernova simulation:
\begin{equation}
    M_{\text{ejecta,i}}=M_{\text{unbound,i}}+M_{\text{envelope,i}}.
\end{equation}
\noindent
Most of these cells lie outside the simulation region. We expect that this mass would eventually be unbound with only minor changes in the composition if the simulation were to be continued to shock breakout. We consider all unstable isotopes to have decayed to stable isotopes. In GCE, we consider elements, not isotopes, so we sum together all isotopes of the same element. During their evolution of hydrostatic burning, the stars of solar metallicity will have ejected some of their mass in stellar winds. The mass of the wind can be obtained from the difference of the mass of the star prior to the supernova and in the ZAMS phase:
\begin{equation}
    M_{\text{wind,i}} = X_{\text{wind,i}}\left(M_{\text{ZAMS}}-M_{\text{preSN}}\right).
\end{equation}
\noindent
The composition of the wind is not provided for the WHW02 progenitors. Compared to the explosive yields, the contribution from the wind to the final yields is negligible for all elements except for hydrogen and helium. We take the wind into account for mass conservation and assume a composition purely of hydrogen and helium. The ratio of those is inferred from wind data for a similar set of progenitors \citep{Woosley2007NucleosynthesisRemnantsMassive}. Only for the most massive progenitors ($30-75\ M_\odot$), this assumption underestimates the wind contribution for carbon and oxygen.

For the initial composition, we multiply the expected mass fractions to the ZAMS mass of the model. For solar metallicity, we use the solar mass fractions of \citet{Anders1989AbundancesElementsMeteoritic}. For sub-solar metallicity, we rescale these mass fractions. For primordial metallicity, we assume the mass fractions $X=0.76$, $Y=0.24$, and $Z=0.00$ for H, He, and the other elements, respectively. The final net yields that are produced (or destroyed if negative) result in
\begin{equation}
    M_{\text{final,i}}=M_{\text{ejecta,i}}+M_{\text{wind,i}}-M_{\text{ZAMS,i}}.
\end{equation}

In Table \ref{tab:net_yields}, we show exemplarily the net yields of some progenitors of solar metallicity corresponding to simulations adopting $\text{HF}=2.7$. The complete set of yields (both total and net) for all simulations are provided as supplementary material online. The total yields also include the amount of the isotopes $^{44}$Ti and $^{56}$Ni that are ejected, because those two isotopes are of particular interest when studying CCSNe.

\begin{table*}
    \centering
    \begin{tabular}{cccccccc}
        \hline
       ZAMS mass & H & He & C & O & Ne & Mg & Si \\
       ($M_\odot$) & ($M_\odot$) & ($10^{-1} M_\odot$) & ($10^{-2} M_\odot$) & ($10^{-1} M_\odot$) & ($10^{-1} M_\odot$) & ($10^{-2} M_\odot$) & ($10^{-2} M_\odot$) \\
       \hline
        $11$ & $-2.20089$ & $3.81529$ & $4.54050$ & $1.70681$ & $0.12199$ & $0.07881$ & $6.94320$ \\
        $12$ & $-7.70578$ & $-30.0603$ & $-3.68734$ & $-1.15572$ & $-0.21069$ & $-0.79318$ & $-0.85413$  \\
        $13$ & $-2.75050$ & $4.57666$ & $5.95173$ & $4.14932E$ & $1.10457$ & $2.60037$ & $5.38514$ \\
        $14$ & $-3.09172$ & $5.04437$ & $7.05349$ & $5.07848$ & $0.85575$ & $4.94468$ & $4.72663$ \\
        $15$ & $-3.43160$ & $5.44857$ & $9.01705$ & $6.50979$ & $1.25604$ & $5.55200$ & $5.72358$ \\
        \hline
        \noalign{\vskip 2mm}  \\
        \hline
          ZAMS mass & S & Ar & Ca & Ti & Cr & Fe & \\
          ($M_\odot$) & ($10^{-2} M_\odot$) & ($10^{-3} M_\odot$) & ($10^{-3} M_\odot$) & ($10^{-4} M_\odot$) & ($10^{-4} M_\odot$) & ($10^{-2} M_\odot$) & \\
        \hline
        $11$ & $4.70651$ & $8.90666$ & $3.95136$ & $0.93519$ & $4.96001$ & $4.26658$ \\
        $12$ & $-0.50203$ & $-1.11522$ & $-0.74475$ & $-0.34951$ & $-2.13347$ & $-1.52919$ \\
        $13$ & $2.82345$ & $4.79914$ & $3.75101$ & $1.16816$ & $6.91629$ & $5.24713$ \\
        $14$ & $2.01478$ & $3.18470$ & $2.73170$ & $1.31465$ & $5.65519$ & $5.07530$ \\
        $15$ & $2.60330$ & $4.27698$ & $3.82888$ & $1.65676$ & $9.56156$ & $8.19982$ \\
        \hline
    \end{tabular}
    \caption{First lines of table 'net\_yields\_hf2.7\_s' in the supplementary material. This table contains the net yields calculated as described in Section \ref{yields} corresponding to the simulations with progenitors of solar metallicity and adopting $\text{HF}=2.7$. For the model with ZAMS mass $12 M_\odot$, all net yields are negative, because this model did not explode. The complete set of yields (both net and total) for all simulations are provided in tables online. All yields (as well as the ZAMS masses) are in units of solar masses.}
    \label{tab:net_yields}
\end{table*}

\section{Galactic chemical evolution} \label{gce}

In this Section, we present the chemical evolution models adopted in order to study the effect of the different massive star yields on the chemical evolution of the solar neighbourhood. In particular, the models are as follows:
\begin{itemize}
    \item the one-infall model (see e.g., \citealp{MatteucciGreggio1986, Matteucci1989, Grisoni2018}), according to which the solar vicinity forms by means of a single episode of primordial gas infall, with a timescale of $\mathrm{\tau\simeq7\ Gyr}$.
    \item the delayed two-infall model of \citet{Spitoni2021} (see also \citealp{Spitoni2019, Palla2020}), which assumes that the Milky Way (MW) disc formed as a result of two different gas infall episodes. The delayed two-infall model is a variation of the classical two-infall model of \citet{Chiappini1997} (see also \citealp{Chiappini2001}) developed in order to fit the dichotomy in $\alpha$-element abundances observed both in the solar vicinity (e.g., \citealp{Hayden2014, recio-blanco2014, Mikolaitis2017}) and at various radii (e.g., \citealp{Hayden2015}). The model assumes that the first primordial gas infall event formed the \textit{chemically} thick disc (namely the high-$\alpha$ sequence) whereas the second infall event, delayed by $\mathrm{\sim 4\ Gyr}$ formed the \textit{chemically} thin disc (the low-$\alpha$ sequence). We stress that the two-infall model adopted here is not aiming to distinguish the thick and thin discs populations geometrically or kinematically (see \citealp{Kawata2016} for a discussion).
\end{itemize}

The models are multi-zones, but here we focus on the evolution of the solar neighborhood only, corresponding to the Galactocentric distance $\mathrm{R_{GC}=8\ kpc}$. We adopt the metallicity dependent stellar lifetimes of \citet{Schaller1992} for stars with $\mathrm{0.1 \leq M/M_\odot \leq 120}$.

\subsection{The model}

The basic equations which describe the evolution of the fraction of gas mass in the form of a generic chemical element $i$ are:
\begin{equation}
    \dot{G}_i(R,t)=-\psi(R,t)X_i(R,t)+\dot{G}_{i,\text{inf}}(R,t)+R_i(R,t),
\end{equation}
where $X_i(R,t)$ represents the abundance by mass of a given element~$i$. $\psi(R,t)X_i(R,t)$ is the rate at which the element $i$ is subtracted from the ISM to be included in stars. $\psi(R,t)$ is the star formation rate (SFR), here parametrized according to the Schmidt-Kennicutt law (\citealp{Kennicutt1998}):
\begin{equation}
    \psi(R,t)\propto\nu \sigma_\text{gas}(R,t)^k,
\end{equation}
with $k=1.5$ and $\sigma_\text{gas}$ being the surface gas density. The star formation efficiency, $\nu$, is a function of the Galactocentric distance, with $\mathrm{\nu(8\ kpc)=1\ \text{Gyr}^{-1}}$. $\dot{G}_{i,\text{inf}}(R,t)$ is the rate at which the element $i$ is accreted through infall of gas. For the two-infall model, the accretion term is computed as:
\begin{equation}
   \dot{G}_{i,\text{inf}}(R,t)=AX_{i,\text{inf}}e^{-\frac{t}{\tau_1}}+\theta(t-t_\text{max})BX_{i,\text{inf}}e^{\frac{t-t_\text{max}}{\tau_2}},
\end{equation}
where $X_{i,\text{inf}}$ is the composition of the infalling gas, here assumed to be primordial for both the infall events. $\tau_1\simeq0.1\ \text{Gyr}$ and $\tau_2\simeq4\ \text{Gyr}$ are the infall timescales for the first and the second accretion event, respectively, and $t_\text{max}\simeq4\ \text{Gyr}$ is the time for the maximum infall on the disc and it corresponds to the start of the second infall episode. The parameters $A$ and $B$ are fixed to reproduce the surface mass density of the MW disc at the present time in the solar neighborhood. In case of single infall, the formulation for the accretion of gas is the following, simpler, one:
\begin{equation}
    \dot{G}_{i,\text{inf}}(R,t)=BX_{i,\text{inf}}e^{-\frac{t}{\tau}}.
\end{equation}
\noindent
Different gas flows than the infall ones (Galactic winds and/or Galactic fountains) are not included. In particular, Galactic fountains, which are more likely to occur in disc galaxies, should not impact in a significant manner the chemical evolution of the disc (see e.g. \citealp{Melioli2009}). 

Finally, $R_i(R,t)$ is the rate of restitution of matter from the stars with different masses into the ISM in the form of the element $i$. This last term takes into account a large variety of stars and phenomena, such as stellar winds, SN explosions of all types, novae and neutron-star mergers. It depends also on the initial mass function (IMF), which here is taken from \citet{Kroupa1993}. In this work, we are interested in testing the yields prescriptions of CCSNe described in the previous Sections. Concerning the stellar yields of the other sources, for all stars sufficiently massive to die in a Hubble time, the following prescriptions have been adopted:\\
\begin{itemize}
    \item for low-intermediate mass stars (LIMS, $\mathrm{1\leq M/M_\odot\leq8}$), we adopt the non-rotational set of yields available on the F.R.U.I.T.Y. data base (\citealp{Cristallo2009, Cristallo2011, Cristallo2015}).
    \item for Type Ia SNe, we assume the single-degenerate scenario for the progenitors with stellar yields from \citet{Iwamoto1999} (their model W7).
    \item for nova systems we adopt prescriptions from \citet{Jose2007}.
    \item for neutron-star mergers, which do not affect significantly the elements treated in this paper but are important for the production of neutron-capture elements, we adopt the prescriptions described in \citet{Molero2023OriginNeutroncaptureElements}. Magneto-rotational SNe are also included in the model, but exclusively for the production of neutron-capture elements.
\end{itemize}

As described in Section \ref{models}, the yields adopted for massive stars are a grid of models in the mass range $\mathrm{11-75\ M_\odot}$ and initial metallicities $\mathrm{Z = 0.0}$, $\mathrm{Z = 10^{-4}\times Z_\odot}$ and $\mathrm{Z=Z_\odot}$. The adopted solar chemical composition is the one of \citet{Anders1989AbundancesElementsMeteoritic}, corresponding to $\mathrm{Z_\odot\simeq0.02}$. For each metallicity, the models are computed for three different HFs: $\mathrm{2.4, 2.7, 5.6}$. The grid of yields corresponding to the progenitor mass range of $\mathrm{11-30\ M_\odot}$ is provided in steps of $\mathrm{1\ M_\odot}$ for the three metallicities and the three HFs, providing great precision in the computation of the chemical evolution. Grids of yields for the ZAMS stars with masses $\mathrm{35}$, $\mathrm{40}$, and $\mathrm{75\ M_\odot}$ are provided only for solar metallicity. For each HF, stars at solar metallicity that are not exploding are assumed to contribute to the chemical enrichment only through the wind. Because of the high precision in the progenitors stellar masses, the chemical evolution model does not require interpolation in large mass ranges for CCSN yields. On the other hand, the model interpolates in the mass range between super-AGB stars and low mass CCSNe (in this model this mass range is $\mathrm{8-11\ M_\odot}$, but it can vary depending on the adopted sets of yields from LIMS and massive stars), where no complete grid is currently available. In fact, the recently new set of yields from \citet{Limongi2024} for this mass range provides yields only at solar metallicity while for super and massive-AGB stars, yields from \citealp{Doherty2014_a, Doherty2014_b} are computed only until $\mathrm{9.0\ M_\odot}$. This interpolation can potentially be a source of uncertainty since, once weighted with the IMF, this mass range is one of the major contributors to the Galactic enrichment (see \citealp{Romano2010QuantifyingUncertaintiesChemical} for a full discussion). Mass conservation is ensured, with massive star remnant (neutron star, black hole or white dwarf) masses being equal to the progenitor mass subtracted by the total ejected mass in the form of the different chemical species.

\begin{figure}
    \includegraphics[width=\columnwidth]{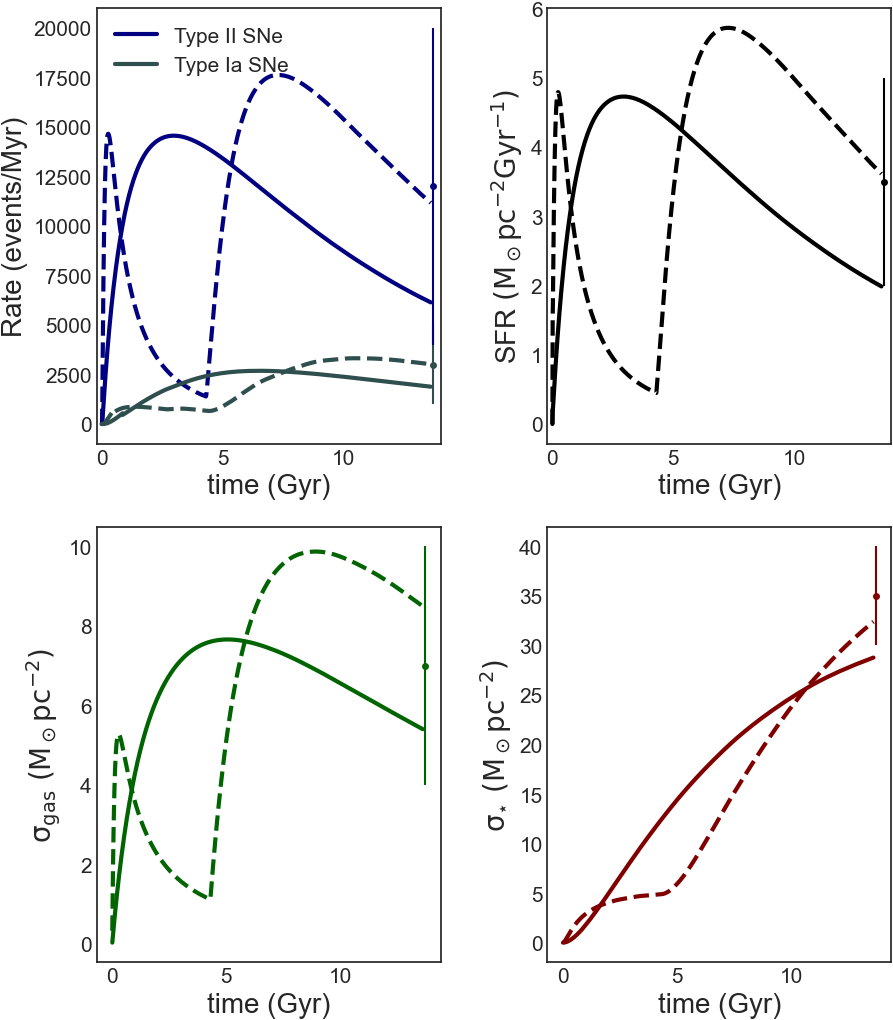}
    \caption{Time evolution of the rate of Type Ia SNe, Type II SNe (upper left panel), SFR (upper right panel), surface densities of gas (lower left panel) and of stars (lower right panel) as predicted by the one- (solid lines) and by the two-infall (dashed lines) models. Predictions of the models are compared to present day values from \citet{Cappellaro1999} (for SNe rates) and from \citet{Prantzos2018} (for SFR and surface gas and stars densities).}
    \label{fig:rates}
\end{figure}

Comparisons of the evolution of some important quantities predicted by our model to present-day observations are reported in Fig.~\ref{fig:rates}. The predicted SFR, and surface densities of stars and gas are computed in the solar neighbourhood and compared with present-day estimates as suggested by \citet{Prantzos2018}. Rates of Type Ia and Type II SNe are averaged over the whole disc and compared with the observational estimates of \citet{Cappellaro1999}. There is a satisfactory agreement with rates of all kinds of SNe as well as with the SFR, and surface densities of gas and stars. In general, the two-infall model provides the best agreement. One of the main observables in the solar neighbourhood is the metallicity distribution function (MDF) which will be discussed in the next Section, because of its dependence on the Fe yield and therefore on the CCSN prescriptions. 

\subsection{Results}

In this Section, we present the results for the solar elemental abundances and MDF predicted by both the one- and the two-infall models. Then, we discuss, for each element, the behaviour of their abundance ratios as functions of metallicity in the typical $\mathrm{[X/Fe]}$ \footnote{the notation $\mathrm{[X/Y]}$ has the meaning $\mathrm{[X/Y]=log(X/Y)-log(X/Y)_\odot}$, with X (Y) being the abundance by number of the element X (Y).} vs. $\mathrm{[Fe/H]}$ diagrams. We will refer to models with HF=2.4, 2.7, 5.6 as models M-24, M-27 and M-56, respectively. For comparison, we show also the results with the non-rotating massive star yields set of \citet{Limongi2018PresupernovaEvolutionExplosive}. We will refer to this model as model M-LC. Because of the differences between our stellar and nucleosynthetic models with those of \citet{Limongi2018PresupernovaEvolutionExplosive}, this comparison is intended to be only qualitative. All the other important parameters of the chemical evolution model, such as IMF, SFR, infall parameters, etc. are kept the same.

\subsubsection{Solar abundances and metallicity distribution function}

\begin{figure}
\includegraphics[width=1\columnwidth]{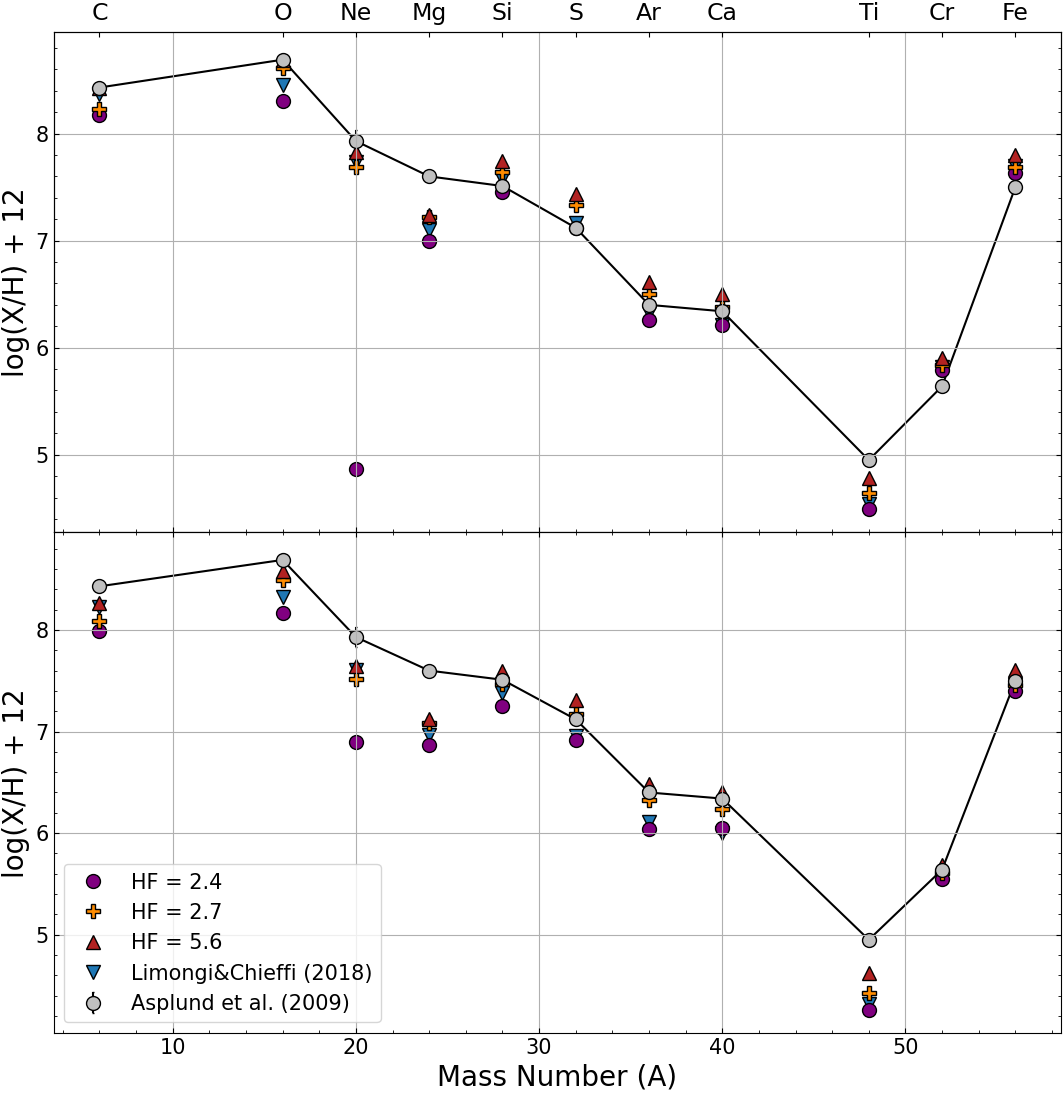}
    \caption{Predicted log(X/H)+12 solar abundances for Fe, C, O, Ne, Mg, Si, S, Ar, Ca, Ti and Cr obtained at the time of the formation of the Solar System by one- (upper panel) and two-infall (lower panel) models using different SNe Type II yield sets. The predictions are compared to the observed photospheric solar abundances by \citet{Asplund2009}.}
    \label{fig: solar}
\end{figure}
\noindent
In Fig.~\ref{fig: solar}, we show the solar abundances for Fe, C, O, Ne, Mg, Si, S, Ar, Ca, Ti and Cr predicted by the one- and the two-infall models, compared with the observed photospheric solar abundances by \citet{Asplund2009}. At solar metallicity, the adoption of a larger HF increases considerably the production of all elements, not necessarily in a linear fashion. For the majority of the elements, the models which better agree with the observed solar abundances are model M-27 and M-56, with either $\mathrm{HF=2.7}$ or $\mathrm{HF=5.6}$. The best fit with the Fe abundance, in particular, is obtained for the two-infall model with $\mathrm{HF=2.7}$. Similar results are obtained by adopting the yields from \citet{Limongi2018PresupernovaEvolutionExplosive}, which here have been used as a second, theoretical, comparison. In particular, since our new CCSN models do not include rotation, we are using the \citet{Limongi2018PresupernovaEvolutionExplosive}'s yields set R with no initial rotational velocity. However, we remind the reader that the inclusion of rotation in stellar evolutionary models (and as a consequence in chemical evolution studies as well) significantly alters the outputs of traditional nucleosynthesis, increasing the yields of almost all elements by factors which depend on the metallicity (we refer to \citealp{Cescutti2013, Prantzos2018, Romano2019, Rizzuti2019, Molero2024} for more discussions on the effect of massive stellar rotators to the chemical evolution of different nuclear species). The majority of the solar abundances for elements studied here are reproduced to more than 95~per cent by the two-infall model with higher HF, with the exception of Ne which is underproduced in almost all models and the \textit{usual} exceptions of Mg and Ti, which are underproduced. The problem of the Mg and Ti underproduction is present in the majority of massive stars yield sets, including \citet{Woosley1995EvolutionExplosionMassive, Francois2004, Limongi2018PresupernovaEvolutionExplosive} and it will be discussed further in Section \ref{sec: abundance ratios}. 

\begin{figure*}
\includegraphics[width=1\textwidth]{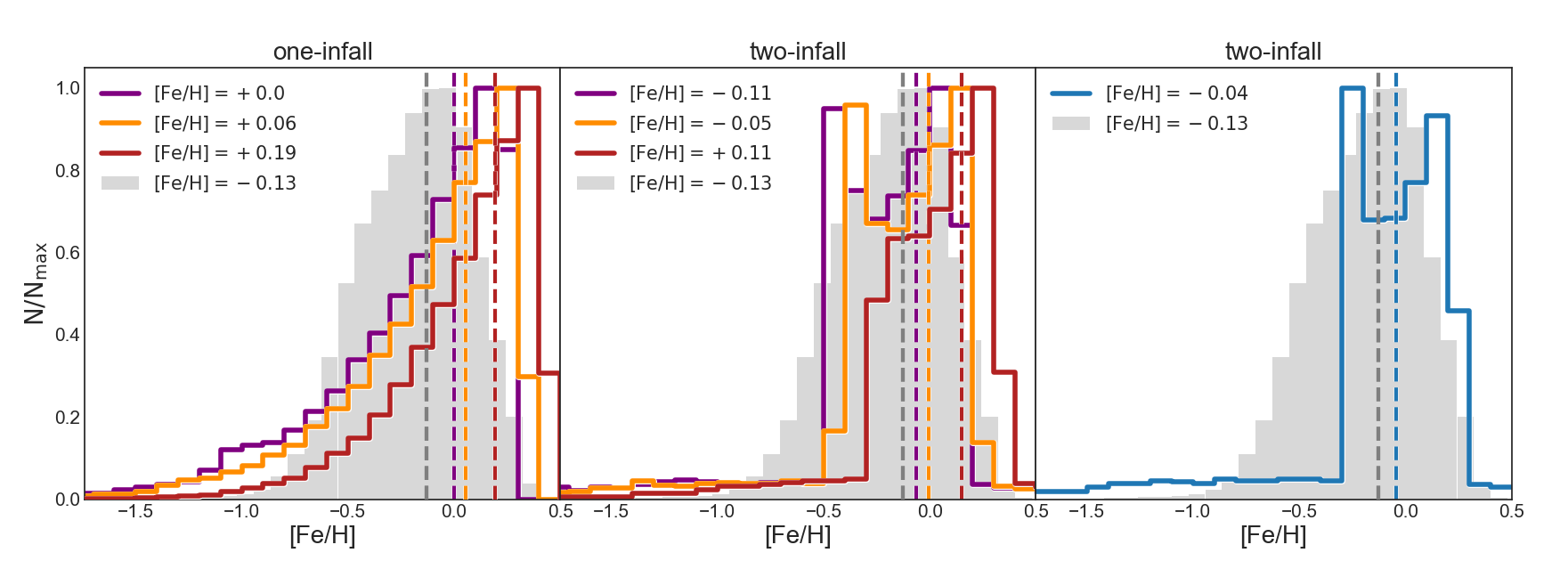}
    \caption{MDF in the solar neighborhood predicted by the one- (left panel) and the two- (middle and right panels) infall models for the three different HFs ($\mathrm{HF=2.4}$ purple line; $\mathrm{HF=2.7}$ orange line; $\mathrm{HF=5.6}$ red line) and for the \citet{Limongi2018PresupernovaEvolutionExplosive}'s yields (blue line) compared with data from the SDSS/APOGEE DR17 sample (grey region). In the legends are reported the predicted and observed values of the median of the distributions.}
    \label{fig: MDFs}
\end{figure*}

In Fig.~\ref{fig: MDFs}, we present our results for the solar neighbourhood MDF predicted by the one- and the two-infall models for the different sets of yields adopted. The predicted MDFs are compared with data from the SDSS/APOGEE DR17 sample of stars with Galactocentric distance $\mathrm{7\leq R_{GC}\leq 9\ kpc}$. The observed MDF is characterized by a single-peak shape, with the peak position at $\mathrm{[Fe/H]\simeq0.0\ dex}$ and a median at $\mathrm{[Fe/H]=-0.13\ dex}$. The single-peak shape is nicely obtained also by our one-infall model, for each value of the HF. The one-infall model results tend, however, to overestimate the number of stars both in low-metallicity ($\mathrm{[Fe/H]\leq-0.75\ dex}$) and in the high-metallicity ($\mathrm{[Fe/H]\geq-0.25\ dex}$) regions, with this latter in a more severe manner. Increasing the HF increases the production of Fe and, as a consequence, causes a shift in the MDF towards higher metallicity. This effect is reduced with the two-infall model, for which the predicted MDFs remain confined in the observed metallicity range even for larger HFs. For $\mathrm{HF=5.6}$, the model still overestimates the number of stars at high [Fe/H], but in a less dramatic way with respect to the correspondent one-infall model. Moreover, for all the HFs, the two-infall models do not overestimate stars in the low-metallicity regime. In the case of the non-rotational set of yields from \citet{Limongi2018PresupernovaEvolutionExplosive}, the two-infall models do not show particular differences. By looking at the overall MDF shape and at the predicted median values (see legends in Fig.~\ref{fig: MDFs}), the models that agree better with the observed MDF, are the two-infall models with low to intermediate HFs (models M-24 and M-27). The MDF shape has a strong dependence on the stellar yields because of the different Fe production. However, it must be reminded that to properly constrain the model with the MDF, dynamical features such as stellar migration and radial gas flows should be taken into account (see e.g., \citealp{Minchev2013, Spitoni2015, Grisoni2018, Palla2020, Vincenzo2020}). Nevertheless, since this work is restricted to the study of the impact of new yield sets on the chemical evolution, it most probably benefits the most in the framework of a simpler GCE model. Therefore, as a first approach, dynamical effects on stars and gas have been excluded.

\subsubsection{Abundance ratios}
\label{sec: abundance ratios}

\begin{figure*}
\begin{center}
    \subfloat{\includegraphics[width=0.9\textwidth]{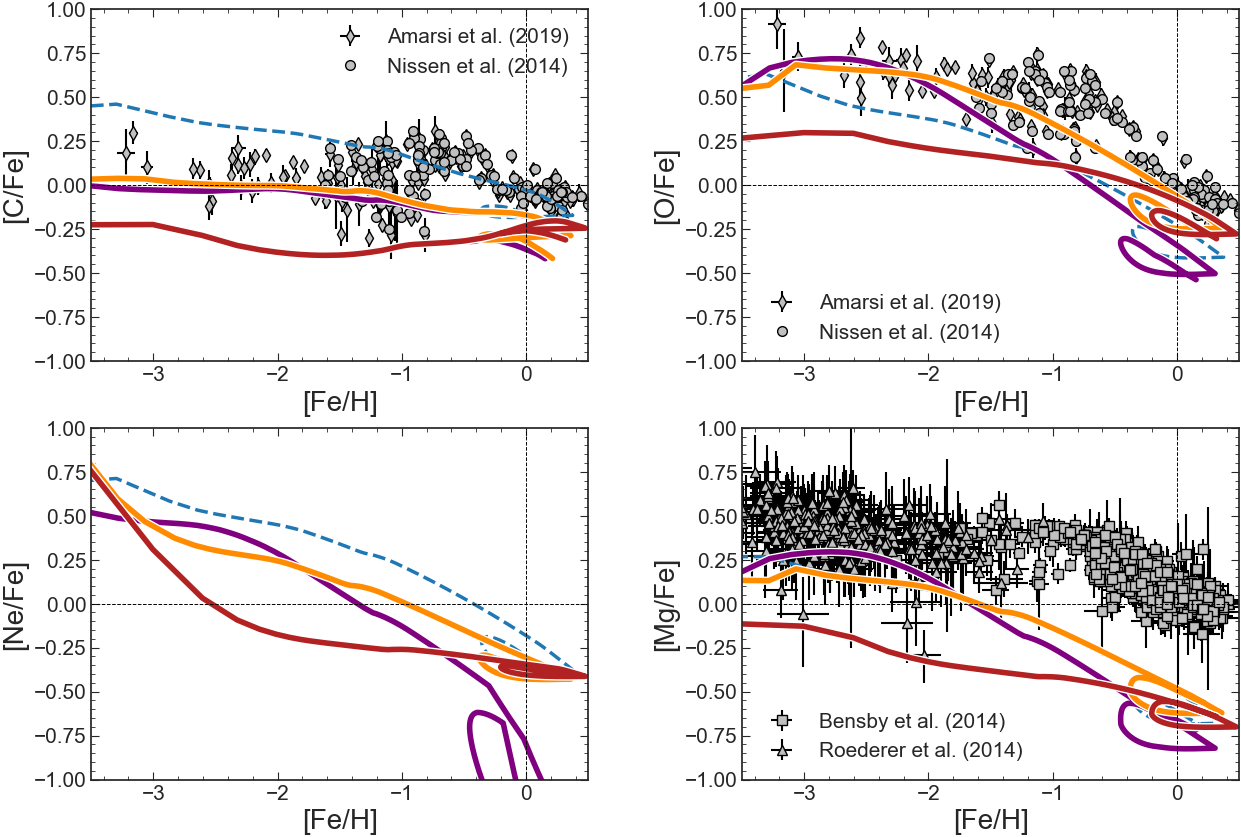}}
    \vfill
    \subfloat{\includegraphics[width=0.9\textwidth]{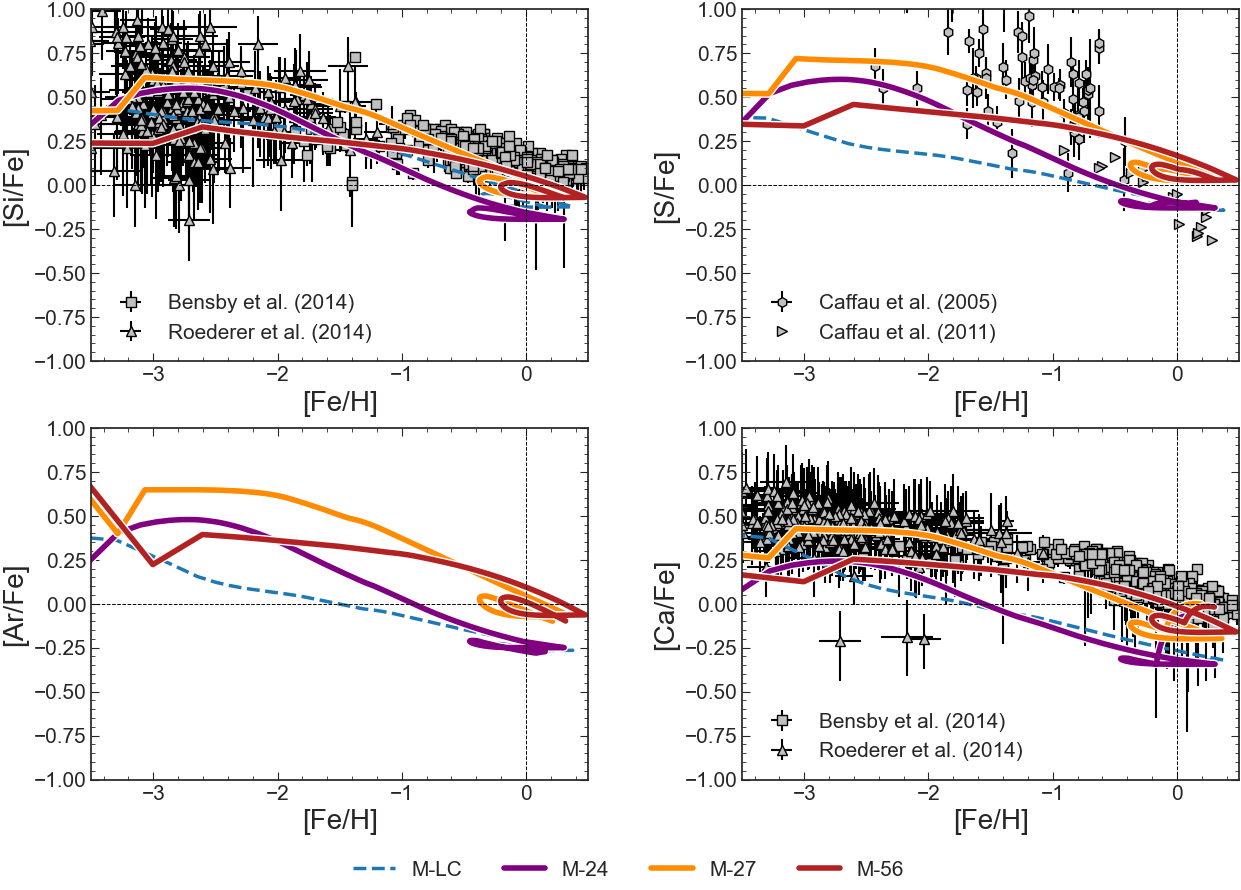}}
%    \vfill
%    \subfloat{\includegraphics[width=1\textwidth]{Chemo_figures/Ti_Cr.png}}
    \caption{Evolution of abundance ratios [X/Fe] vs. [Fe/H] for C, O, Ne, Mg, Si, S, Ar and Ca compared to observational data. Model M-24, M-27 and M-56 include the new sets of yields with HFs $\mathrm{HF=2.4, 2.7}$ and $\mathrm{5.6}$, respectively. Model M-LC correspond to the non-rotational yield set R from \citet{Limongi2018PresupernovaEvolutionExplosive}}
    \label{fig: abundances 1,2}
\end{center}
\end{figure*}

\begin{figure*}
\includegraphics[width=0.9\textwidth]{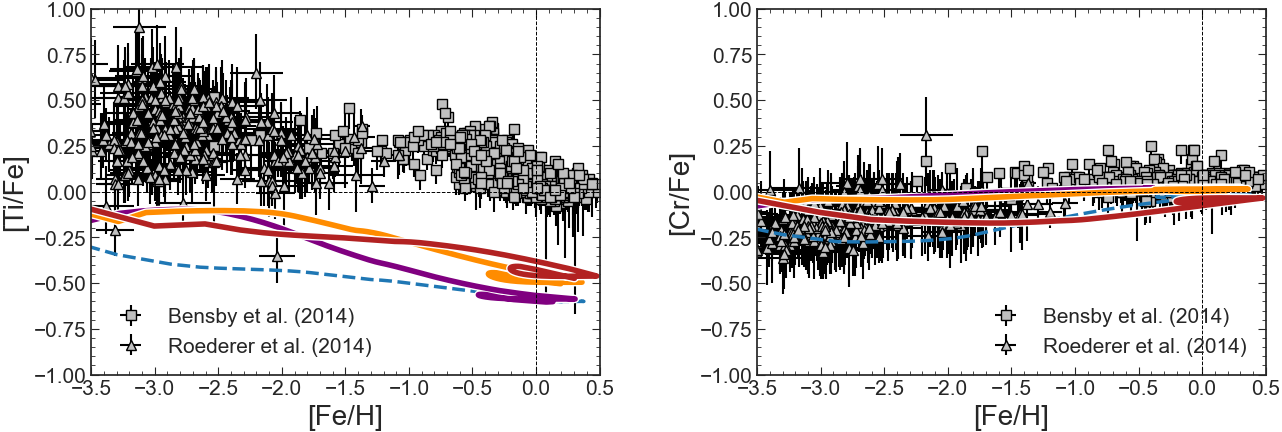}
    \caption{Same as Fig.~\ref{fig: abundances 1,2} but for Ti and Cr.}
    \label{fig: abundances 3}
\end{figure*}

In this Section, we present the predicted behaviour of the different abundance ratios as a function of metallicity for the elements produced by the CCSN models. We use Fe as a metallicity indicator and show the results in the typical [X/Fe] vs. [Fe/H] diagnostic diagrams. We will show results only for the two-infall models, which, with the current set of yields, agrees better with the solar abundances as well as with the MDF, as shown in the previous Section. Results are displayed in Figures \ref{fig: abundances 1,2} and \ref{fig: abundances 3}. We compare our model results with data from only a few surveys, to keep the data set as homogeneous as possible and to not overcrowd the plots. Nevertheless, some dispersion in the data is still visible at low metallicity probably due to the inhomogeneities in the ISM (see e.g., \citealp{Cescutti2008} and \citealp{Cescutti2014}).
\\
\\
\textsc{Carbon}: For the analysis of the [C/Fe] vs. [Fe/H] trend, we adopt data from \citet{Nissen2014} and \citet{Amarsi2019} (grey circles and diamonds in Fig.~\ref{fig: abundances 1,2}, respectively). Note that this is the same observational set adopted by \citet{Romano2019}, where the evolution of the CNO isotopes is widely investigated (see also \citealp{Romano2003, Romano2017}). \citet{Nissen2014} determined abundances for C, O and Fe for F and G main-sequence stars in the solar vicinity with $\mathrm{-1.6<[Fe/H]<+0.4\ dex}$. Carbon abundances are determined from the $\lambda \lambda 5052, 5380\ \rm{CI}$ lines to which 1D non-local thermodynamic equilibrium (non-LTE) corrections were applied. The observed stars belong to different stellar populations, including both thin and thick discs as well as the halo. \citet{Amarsi2019} carried out 3D non-LTE radiative transfer calculations for $\rm CI$ and $\rm OI$. The derived absolute corrections between the 3D non-LTE and 1D LTE were used in the re-analysis of C, O and Fe in F and G dwarfs belonging to the MW disc and halo with $\mathrm{[Fe/H]\leq-1.5\ dex}$. The behaviour of [C/Fe] at low metallicities observed in the data from \citet{Amarsi2019} is rather flat, with almost a solar value. This flat trend is nicely reproduced by models M-24 and M-27 that assume low HFs (2.4 and 2.7). The two models predict similar results for all the range of metallicity, with model M-27 only slightly above model M-24. The increase of the HF to the high value of $\mathrm{HF=5.6}$ does not translate into a higher [C/Fe] trend. Since, after weighting by the IMF, the Fe enrichment is greater than the C one, the line corresponding to model M-56 lies below those of models M-24 and M-27 for all the range of metallicity, reaching similar values only at solar. The $\mathrm{[C/Fe]\sim0}$ produced by model M-24 and M-27 is obtained also with the yields from \citet{Woosley1995EvolutionExplosionMassive} and from \citet{Nomoto2013} (without hypernovae). Data with $\mathrm{[Fe/H]\geq-1.6}$ from \citet{Nissen2014} are not reproduced by the models with our new sets of yields, which underestimate the [C/Fe] ratio. On the other hand, model M-LC provides an adequate fit to this observed trend, but it fails to reproduce the flat trend at low [Fe/H].
\\
\\
\textsc{$\alpha$-elements}: For the $\alpha$-elements O, Si, S, and Ca, we adopt different sets of observational data from \citet{Caffau2005, Caffau2011, Nissen2014, Bensby2014, Roederer2014} and \citet{Amarsi2019}. \citet{Bensby2014} conducted a high-resolution spectroscopic study of F-G dwarf and subgiant stars in the solar neighborhood. In particular, they selected stars kinematically belonging to the thin and thick discs, the metal-rich halo as well as to sub-structures as the Hercules stream and the Arcturus moving group. The elemental abundances have been determined using 1D, plane-parallel model atmospheres under the LTE assumption, correcting the $\rm{FeI}$ lines for non-LTE effects. The low metallicity data are from \citet{Roederer2014}. The authors presented detailed abundances from high resolution optical spectroscopy of 313 metal-poor stars. They performed a standard LTE abundance analysis by using 1D model atmospheres for their analysis, applying line-by-line statistical corrections. The observational data adopted for the [S/Fe] vs. [Fe/H] are from \citet{Caffau2005} and \citet{Caffau2011}. \citet{Caffau2005} assembles a sample of 253 stars with metallicity in the range $\mathrm{-3.2 \leq [Fe/H] \leq +0.5\ dex}$ made of both literature data and their own measurements of 74 Galactic stars high resolution spectra. From their sample a significant scatter in the [S/Fe] at $\mathrm{[Fe/H]\simeq-1.0\ dex}$ is observed, that the authors did not attribute necessarily to distinct stellar populations. As shown in Fig.~\ref{fig: abundances 1,2}, data are showing the typical abundance pattern of $\alpha$-elements, characterized by a plateau and by a high [$\alpha$/Fe] ratio at low [Fe/H] values, where the production of both $\alpha$-elements and Fe is only due to CCSNe, and a decrease in [$\alpha$/Fe] ratio with higher [Fe/H] because of the late injection of Fe from Type Ia SNe (\textit{time-delay model}, \citealp{Matteucci2012}). Model M-24, with the lowest HF, tends to underestimate the observed abundance pattern of all $\alpha$-elements and for almost all the range of metallicity. In fact, for $\mathrm{HF=2.4}$, only few progenitors are actually exploding at primordial and solar metallicity. The majority of them are exploding at intermediate metallicity, causing the peak in the [$\alpha$/Fe] at $\mathrm{[Fe/H]\simeq-3\ dex}$. After the peak, the model quickly decreases with metallicity because of the strong contribution to the Fe production by Type Ia SNe, which is not being compensated by CCSNe. As a consequence, model M-24 cannot reproduce the typical 'plateau' like pattern of $\alpha$-elements, and, except at $\mathrm{[Fe/H]\leq-2\ dex}$ for O and Si, it underestimates the general trend for the entire range of metallicity. Increasing the HF to the high value of $\mathrm{HF=5.6}$ allow us to obtain results in agreement with the typical 'plateau' like trend but, because of the stronger production of Fe than $\alpha$-elements, the model lies under the observed trend for all the range of metallicity, similarly to what happens with C. The underproduction is not critical in the case of Si, S and Ca, in particular at intermediate and at solar metallicity. In fact, even though the predicted trend is generally below the observed one, some stars have abundances that may be consistent with a stronger explosion, similarly to what has been found by \citet{Romano2010QuantifyingUncertaintiesChemical} in the case of hypernovae. Model M-27, with an intermediate value for the HF, produces the best agreement with the observations for all the $\alpha$-elements, in particular below $\mathrm{[Fe/H]\simeq-1.0\ dex}$. Towards higher [Fe/H], the model tends to slightly underestimate the observed trend, as in the case of O and Ca. However, since this discrepancy is present in a metallicity range where the contribution to the Fe (and partially also to $\alpha$-elements) from Type Ia SNe is not negligible, it is difficult to state to what extend this inconsistency is due to CCSNe only.
\\
\\
\textsc{Magnesium and Titanium}: Concerning Mg and Ti, their evolution is not well reproduced by our model. The solar abundance of these two elements is underestimated, as explained above, and the underproduction characterizes their whole evolution. Only model M-24 provide a slight agreement with the data from \citet{Roederer2014} for the [Mg/Fe] at low [Fe/H] values, but it is considered by no means to be satisfactory. Depending on the model, [Mg/Fe] and [Ti/Fe] are $\sim0.25-0.4\ \mathrm{dex}$ underabundant overall. The underproduction of Mg and Ti is a well known problem in GCE. Similar results to those obtained here arise also with yields of different authors (as for example those of \citealp{Woosley1995EvolutionExplosionMassive} and of \citealp{Limongi2018PresupernovaEvolutionExplosive} for which we refer to the corresponding chemical evolution studies of \citealp{Timmes1995} and \citealp{Prantzos2018}, respectively), which often need to be increased by large factors (see e.g. \citealp{Francois2004}). An alternative solution for Mg can be represented by the inclusion of hypernovae in the chemical evolution calculations, as shown by \citet{kobayashi2006} and by \citet{Romano2010QuantifyingUncertaintiesChemical}. In both cases, $\sim 50$~per cent of stars more massive than $20\ M_\odot$ should explode as hypernovae (this fraction can slightly change depending on the adopted IMF). On the other hand, up to date, no solution is known to correct the Ti abundances (see however \citealp{Maeda2003}). We remind the reader that this is not related to the problem of producing sufficient $^{44}$Ti in 1D CCSN simulations (see e.g. \citealp{Wang2024InsightsProduction44Ti} for a discussion) as this radioactive isotope decays to $^{44}$Ca, while our Ti comes from the decay of $^{48}$Cr.
\\
\\
\textsc{Noble gases}: We report the evolution also of the noble gases Ne and Ar for which, however, there are no observations in stars. Ne and Ar are $\alpha$-elements with plateau values predicted by our models of $\mathrm{[Ar-Ne/Fe]\sim0.5\ dex}$ for $\mathrm{[Fe/H]\leq-1.0\ dex}$. Indeed, increasing the HF produces similar effects in Ne and Ar as in the other $\alpha$-elements. With respect to model M-LC, with \citet{Limongi2018PresupernovaEvolutionExplosive}'s yields, our models tend to underestimate and overestimate the [Ne/Fe] and [Ar/Fe] vs. [Fe/H] trends overall, respectively, independently on the HF. 
\\
\\
\textsc{Chromium}: Model M-24 and M-27 produce near-solar [Cr/Fe] for all the range of metallicities, with a slightly decreasing trend toward lower [Fe/H]. This is well in agreement with observations which down to $\mathrm{[Fe/H]\sim -1.5\ dex}$ show a constant [Cr/Fe] ratio which then slightly decreases with decreasing metallicity. Since Cr, together with the other Fe-peak elements, is half produced by Type Ia SNe, which contribution is not a function of the metallicity, the increase of the [Cr/Fe] with [Fe/H] obtained by our model is only due to massive stars. Model M-56, tends to underestimate the overall trend, in particular at intermediate [Fe/H], most probably due to the stronger Fe production.

\section{Conclusions}
\label{sec: conclusions}

In this work, we provide a new set of CCSN yields and test them in a consistent model of GCE able to reproduce the main features of the Milky Way disc and of the solar vicinity. The new grid of stellar yields covers the mass range $\mathrm{11-75\ M_\odot}$, mostly in steps of $\mathrm{1\ M_\odot}$, and the initial metallicities $\mathrm{Z=0.0, 10^{-4}, 1.0\ Z_\odot}$. In our spherically symmetric, neutrino-driven CCSN simulations, additional energy is injected with a 'heating factor' (HF) to mimic the support to the neutrino heating mechanism by convection and turbulence. We test the impact of different calibration methods for these artificial explosions and find that the results significantly change when calibrating to SN1987A or to a 3D simulation. This uncertainty estimate is propagated by using  three sets of yields with different HFs (2.4, 2.7, and 5.6) in our GCE model. The uncertainty in the CCSN yields is still relevant after weighting by the IMF and combining with other stellar sources. All predictions from our GCE model differ significantly for our three sets of yields. Generally, a higher HF increases the production of all elements. Interestingly, this does not result in linear trends in the abundance ratios.

The model that best reproduces the observations is the one with $\text{HF}=2.7$. It has the best agreement with the shape of the solar MDF, as well as the solar iron abundance, in particular in the case of a two-infall scenario of disc formation. This model also nicely reproduces the order of magnitude of the $\alpha$-elements abundance ratios as well as the trend with metallicity. The [Mg/Fe] and [Ti/Fe] vs. [Fe/H] abundance trends are underproduced with a slight improvement with respect to the non-rotational set of \citet{Limongi2018PresupernovaEvolutionExplosive}. The evolution of [Cr/Fe] with [Fe/H] as well as the plateau observed in the [C/Fe] at low [Fe/H] is well reproduced by both models with lower HF.

Comparing the sets with $\text{HF}=2.4$ and $\text{HF}=2.7$, the main difference is that only about half of the stars explode with the lower HF. The question of the explodability pattern (which stars explode and how many of them) remains open. Our simulations, though 1D, reproduce the explodability pattern of recent 3D simulations. How many and which stars explode significantly changes the final yield predictions, because for a star that does not explode, the CCSN contribution is lost. Since the set with the lowest HF underproduces almost all elements, this suggests that at least half of the massive stars have to explode to fit the observed chemical evolution of the solar neighbourhood.

We find a positive correlation between the amount of iron that is ejected and the explosion energy. This has been found in previous studies and shows that the dynamics of the explosions have a direct impact on the nucleosynthesis. Also the yields of the other elements are increased with higher explosion energy, but they do not scale as strongly. This is the main reason for the differences between the models with $\text{HF}=2.7$ and $5.6$. Most models explode in both cases, but, in general, the explosions with the higher HF are more energetic. The increased iron production reduces the ratio with respect to iron ([X/Fe]) for all elements and thus their abundance patterns seem to be underproduced. As a result, increased explosion energies of about $1-\SI{2.5}{\bethe}$ lessen the agreement with the observational data. This indicates that the typical CCSN explosion energy should be lower than about $\SI{2}{\bethe}$.

With this work, we proved the great importance of building a self-consistent chain between the final stage of stellar evolution, nucleosynthesis calculations and GCE models. We conclude that using late advances in spherically symmetric CCSN simulations or even the growing set of multi-dimensional simulations to obtain yields for GCE models can provide useful constraints to both fields.

\section*{Acknowledgements}

We thank the referee Friedrich-Karl Thielemann for the useful comments that improved the manuscript. We thank Moritz Reichert and Benedikt Riemenschneider for useful discussions. This work was supported by the Deutsche Forschungsgemeinschaft (DFG, German Research Foundation) -- Project-ID 279384907 - SFB 1245 and by the State of Hesse within the Research Cluster ELEMENTS (Project ID 500/10.006). The authors gratefully acknowledge the computing time provided to them
on the high-performance computer Lichtenberg at the NHR Centers NHR4CES at TU Darmstadt (Project-ID 2055). This is funded by the Federal Ministry of Education and Research, and the state governments participating. Furthermore, the authors acknowledge the support through the grant PID2021-127495NB-I00 funded by MCIN/AEI/10.13039/501100011033 and by the European Union, the Astrophysics and High Energy Physics programme of the Generalitat Valenciana ASFAE/2022/026 funded by MCIN and the European Union NextGenerationEU (PRTR-C17.I1), and the Prometeo Excellence Programme (grant CIPROM/2022/13) funded by the Generalitat Valenciana.  MO was supported by the Ramón y Cajal programme of the Agencia Estatal de Investigación (RYC2018-024938-I). M. Molero and F. Matteucci thank I.N.A.F. for the 1.05.12.06.05 Theory Grant - Galactic archaeology with radioactive and stable nuclei.

%%%%%%%%%%%%%%%%%%%%%%%%%%%%%%%%%%%%%%%%%%%%%%%%%%
\section*{Data Availability}

The tables with the complete set of CCSN yields (both total and net) can be found as extra material available online.



\bibliographystyle{mnras}
\bibliography{YieldsGCE} % if your bibtex file is called example.bib

% Don't change these lines
\bsp	% typesetting comment
\label{lastpage}
\end{document}